\documentclass[a4paper,11pt]{article}
\pdfoutput=1

\usepackage{jheppub}
\usepackage[utf8]{inputenc}
\usepackage{amsmath, amssymb}
\usepackage{mathrsfs}
\usepackage{float}
\usepackage{color}
\usepackage{xcolor}
\usepackage{multirow}
\usepackage{graphicx,caption,subcaption}
\usepackage{bm}
\usepackage{dsfont}
\usepackage{cancel}
\usepackage{latexsym}
\usepackage{slashed}
\usepackage[space]{grffile}
\usepackage{tikz}
\usepackage{enumitem}
\usepackage{upgreek}

\newcommand{\Ga}{{\Gamma}}
\newcommand{\De}{{\Delta}}
\newcommand{\Lm}{{\Lambda}}

\newcommand{\al}{{\alpha}}
\newcommand{\bt}{{\beta}}
\newcommand{\ga}{{\gamma}}
\newcommand{\de}{{\delta}}
\newcommand{\ep}{{\epsilon}}
\newcommand{\vep}{{\varepsilon}}

\newcommand{\te}{{\theta}}

\newcommand{\lm}{{\lambda}}

\newcommand{\sig}{{\sigma}}

\newcommand{\mrm}[1]{{\mathrm{#1}}}

\newcommand{\bsN}{{\boldsymbol{N}}}

\newcommand{\bsv}{{\boldsymbol{v}}}

\newcommand{\bscN}{{\boldsymbol{\mathcal{N}}}}

\newcommand{\cC}{{\mathcal{C}}}

\newcommand{\cL}{{\mathcal{L}}}

\newcommand{\cO}{{\mathcal{O}}}

\newcommand{\cR}{{\mathcal{R}}}

\newcommand{\cY}{{\mathcal{Y}}}

\newcommand{\sfD}{{\mathsf{D}}}

\newcommand{\sfL}{{\mathsf{L}}}
\newcommand{\sfM}{{\mathsf{M}}}
\newcommand{\sfR}{{\mathsf{R}}}

\newcommand{\rL}{{\mathrm{L}}}
\newcommand{\rR}{{\mathrm{R}}}

\newcommand{\dg}{{\dagger}}
\newcommand{\mcd}{{\kern-0.1pt\cdot\kern-0.1pt}}

\newcommand{\nn}{{\nonumber}}

\newcommand{\ol}{\overline}
\newcommand{\pd}{{\partial}}

\newcommand{\bbra}{{\langle\kern-2.5pt\langle}}
\newcommand{\kket}{{\rangle\kern-2.5pt\rangle}}
\newcommand{\Bbra}{{\Big\langle\kern-3.5pt\Big\langle}}
\newcommand{\Kket}{{\Big\rangle\kern-3.5pt\Big\rangle}}

\DeclareMathOperator*{\msum}{\mbox{$\sum$}}

\def\simlt{\mathrel{\lower2.5pt\vbox{\lineskip=0pt\baselineskip=0pt
           \hbox{$<$}\hbox{$\sim$}}}}
\def\simgt{\mathrel{\lower2.5pt\vbox{\lineskip=0pt\baselineskip=0pt
           \hbox{$>$}\hbox{$\sim$}}}}

\title{Searching for a Dark Dimension Right-handed Neutrino in KATRIN}

\author[a,b]{Ignatios Antoniadis}
\author[a]{Auttakit Chatrabhuti}
\author[a]{Hiroshi Isono}

\affiliation[a]{High Energy Physics Research Unit, Faculty of Science,
Chulalongkorn University,\\
Bangkok 10330, Thailand}
\affiliation[b]{Laboratoire de Physique Th\'eorique et Hautes Energies - LPTHE\\
Sorbonne Universit\'e, CNRS, 4 Place Jussieu, 75005 Paris, France}

\emailAdd{antoniad@lpthe.jussieu.fr, auttakit.c@chula.ac.th, hiroshi.i@chula.ac.th}

\abstract{We study the possibility that the Right-handed neutrino is a five-dimensional state propagating along a micron size extra dimension, as required in the dark dimension proposal. We work out the signatures of R-neutrino production in KATRIN experiment and compare them with those of a sterile neutrino which manifests by a kink in the electron energy spectrum of the beta-decay at a value corresponding to the sterile neutrino mass. 
We explore the allowed parameter space of the compactification scale and the R-neutrino bulk mass versus the Yukawa coupling, and show that a large part of it is within KATRIN's sensitivity. When the bulk mass is much smaller than the compactification scale, several kinks could be observed corresponding to the positions of the R-neutrino Kaluza-Klein excitations, while for large bulk mass there will be effectively one kink at the position of the bulk mass.}

\begin{document}
\maketitle

\section{Introduction}

Extra dimensions turn out to be an important ingredient of theories of quantum gravity, such as superstring theory~\cite{GSW}. If their size is large compared to the four-dimensional (4D) Planck scale $M_p$, it allows to address the origin of several mass hierarchies in high energy physics, such as those associated with the supersymmetry breaking scale~\cite{Antoniadis:1990ew}, the electroweak scale~\cite{Arkani-Hamed:1998jmv, Antoniadis:1998ig}, the smallness of neutrino masses~\cite{Dienes:1998sb, Arkani-Hamed:1998wuz, Dvali:1999cn} and the scale of dark energy~\cite{Burgess:2004yq, Montero:2022prj}. A crucial observation is that gravity becomes strong at energies lower than $M_p$, of the order of $M_*=M_p/\sqrt{N}$, where $N$ is the number of light degrees of freedom below $M_*$. This is typically the higher dimensional Planck scale, or the string scale, or more generally the so-called species scale~\cite{Dvali:2007hz, Dvali:2007wp}. A model independent experimental bound on the size $R$ of extra dimensions comes from short distance measurements of Newton's gravitational inverse-square law in tabletop experiments, leading to $R\lesssim 30\,\mu$m~\cite{Lee:2020zjt}. For such large sizes, with the Standard Model of particle physics and our observable universe should be localised in the large extra dimensions on an effective three-dimensional brane leaving only gravity propagating in the higher-dimensional bulk~\cite{Arkani-Hamed:1998jmv, Antoniadis:1998ig}.

More recently, an independent theoretical argument on (large) extra dimensions came from the distance conjecture of the swampland program which postulates that a tower of light states should appear at large distances in the landscape of string theory vacua at a mass scale exponentially small in the proper distance with an exponent which of order unity in 4D Planck units~\cite{Vafa:2005ui, Ooguri:2006in}. The decompactification limit is a particular example since the proper distance is proportional to the logarithm of the volume of the compactified space and the tower of light states are identified with the Kaluza-Klein (KK) excitations of the graviton. On the other hand, it was argued that a similar result applies for anti-de Sitter (AdS) vacua with the proper distance being proportional to the logarithm of the cosmological constant $|\Lambda|$~\cite{Lust:2019zwm}. Assuming that this is also valid for (approximate) de Sitter (dS) vacua with positive $\Lambda$ and applying it to the present dark energy, one concludes that a tower of light states should appear in the limit of vanishing $\Lambda$ at a scale proportional to a positive power $\Lambda^\alpha$. Combining theoretical constraints with experimental results, one finds that $\alpha=1/4$ and a KK tower should open up at a scale $R^{-1}$ with
\begin{align}
\label{Rdef}
R=\lambda\,\Lambda_\text{DE}^{-1/4}\simeq{\cal O}(\mu\text{m})\,,
\end{align}
where the proportionality factor is estimated to be within the range $10^{-4} < \lambda < 10^{-1}$. 

Thus, extra dimensions should open up at the micron region associated to the dark energy scale. For one dark dimension $(d=1)$ the species scale $M_*$ (5D Planck mass) is of order $10^9$ GeV, while the case of $d=2$ implies $M_*\sim{\cal O}(10)$ TeV and is at the border of astrophysical and cosmological constraints~\cite{Anchordoqui:2025nmb}.
The Dark Dimension (DD) scenario has several interesting implications for particle physics and cosmology, for a review see for instance~\cite{Anchordoqui:2024ajk} and references therein. In particular, micron size extra dimensions offer a framework that can explain the smallness of neutrino masses by postulating that the Right-handed neutrino propagates in the dark dimension, besides the graviton~\cite{Dienes:1998sb, Arkani-Hamed:1998wuz, Dvali:1999cn, Anchordoqui:2022svl}. 

Indeed, the standard sea-saw mechanism on the brane requires a 4D Majorana state with a mass around $10^{13}-10^{14}$ GeV which is much heavier than the 5D Planck mass, unless there is an extra unnatural coupling suppression by several orders of magnitude. On the other hand, if the R-neutrino $\nu_R$ propagates in the bulk, its coupling to the brane with the localised Left-handed lepton doublet and the electroweak Higgs  (in $M_*$ units), acquires a natural wave function suppression by 9 orders of magnitude $(M_p/M_*)$ which can easily lead to the required bold part of neutrino masses. Besides explaining the neutrino mass hierarchy, this proposal predicts that neutrino masses are of Dirac type excluding neutrinoless double beta-decay which has not yet been observed. Dirac neutrinos are also supported by the swampland AdS conjecture~\cite{Ibanez:2017kvh, Anchordoqui:2023wkm} that forbids stable non-supersymmetric AdS vacua~\cite{Ooguri:2016pdq} which the Standard Model would acquire upon compactification in lower dimensions if neutrinos were Majorana~\cite{Arkani-Hamed:2007ryu}.

In the absence of bulk masses for the R-neutrinos, oscillation data restrict the compactification radius to be less than about 0.4 or 0.2 $\mu$m, for Normal or Inverted Hierarchy, NH or IH respectively~\cite{Machado:2011jt, Forero:2022skg}.\footnote{See also footnote 6 of Ref.~\cite{Anchordoqui:2024ajk}.} 
However, most of these studies focused mainly on the parameter region where the Yukawa coupling is small or of the same order as the compactification scale. Thus, their conclusions may not be valid for large Yukawa coupling limit, relaxing the bound, as we show below in this paper.
In the presence of bulk neutrino masses, the KK spectrum is modified drastically and the bounds on to the compactification radius from neutrino data are essentially relaxed~\cite{Lukas:2000wn, Lukas:2000rg, Carena:2017qhd, Anchordoqui:2023wkm, Eller:2025lsh}; in particular the mass of the zero-mode is suppressed exponentially for large positive bulk mass~\cite{Carena:2017qhd}, while the couplings are picked around a higher KK mode away from the zero-mode~\cite{Anchordoqui:2023wkm}. 

In this work we make a detailed analysis of neutrino physics in the context of the dark dimension with focus on experimental signatures in KATRIN which aims to measure neutrino masses and search for possible sterile neutrinos in the range of order 0.1-100 eV~\cite{KATRIN:2022ith, KATRIN:2024odq, KATRIN:2025lph}. The main signal is that the production of a sterile neutrino leads to a kink in the differential beta decay spectrum as a function of the electron energy near the endpoint $E_0\simeq 18.57$ keV, that could be observed if the mass is within the experimental sensitivity of the energy resolution.
On the other hand, if $\nu_R$ propagate in the bulk of the dark dimension, its KK excitations act as a tower of sterile neutrinos with couplings stemming from its 5D coupling to the brane. In general, there are 3 parameters: the size of the dark dimension $R$ related to the dark energy by \eqref{Rdef}, the bulk mass $c$ of the R-neutrino and its 5D coupling to the brane $\mu$. Here, we find two interesting regions of the parameter space, where analytic results can be obtained, leading to distinct experimental signatures within the reach of KATRIN:
\begin{itemize}
\item When the bulk mass is small compared to the compactification scale, it acts as a small perturbation to the well studied in the past massless case. KK excitations of the R-neutrino would then lead to a series of kinks in the electron energy spectrum that as we show could be within the KATRIN sensitivity for a size of the dark dimension smaller than around the micron region\footnote{Ref.~\cite{Rodejohann:2014eka} considered the model with massless bulk neutrinos and demonstrated a few kinks from KK modes of eV and keV respectively.}. This region includes at its boundary the large bulk R-neutrino to brane coupling $\mu R$ for which previously obtained bounds from oscillations do not apply because the coupling of KK excitations to the lowest active neutrino vanishes in this limit.
\item When the bulk mass is sufficiently large compared to the compactification scale, there is a quasi-continuum of KK excitations above the bulk mass that acts as a mass gap above a very light active neutrino zero-mode. Moreover, their coupling to the zero-mode is picked around the $cR$-th excitation with mass of order the bulk mass. As a result, the experimental signal is effectively one kink in the electron beta-decay energy spectrum around the bulk mass which ressembles the one of an ordinary sterile neutrino, so called 3+1 model, with the same mass and an effective mixing angle which is calculable analytically. We can then provide already a plot of the excluding region in the parameter space, based on existing KATRIN's results.
\end{itemize}

The outline of our paper is the following. In Section~2, we review the more general model describing the mechanism of generating neutrino masses by introducing three R-neutrinos (one for each generation) in the bulk of an extra dimension compactified on a line interval with the Standard Model localised on a 3-brane at one of its ends. Subsection~2.1 presents the action and its KK mode expansion in terms of normalised eigenfunctions with eigenvalues given as solutions of (hyper)trigonometric equation, as well as their relation to the active neutrino flavour basis. In Section~3, we first solve the eigenvalue equation for the KK mass spectrum (subsection~3.1) and we impose the experimental constraints on the mass square differences from neutrino oscillations (subsection~3.2). We then impose all remaining phenomenological constraints on the model parameters and obtain bounds on the compactification scale (subsection~3.3); in particular, we analyse the large $\mu R$ limit for zero bulk mass which leads to a new allowed window for the size of the dark dimension around the micron scale. In Section~4, we apply the above results to study the experimental signatures in KATRIN based on the differential beta-decay spectrum. In particular, we identify two regions of the parameter space where analytic results can be obtained, corresponding to small or large bulk masses compared to the compactification scale (subsection~4.1) and can make an explicit comparison with the experimental results for the 3+1 sterile neutrino model (subsection~4.2), in the case of normal hierarchy. The case of inverse hierarchy is discussed in subsection~4.3. Our conclusions are presented in Section~5. Finally, Appendix~A contains technical details used in the derivation of the KK spectrum.

\section{Setup}
\subsection{Summary of our model}

Our model of neutrino masses originates from a five-dimensional model with three massive fermions (5D spinors) of masses $c_1,c_2,c_3$ which are singlet under the Standard Model (SM) gauge transformations. 
The spacetime is the product of 4D Minkowski spacetime $(x^0,x^1,x^2,x^3)$ and a compact direction $z\in[0,\pi R]$ (line segment). 
At the end point $z=0$ is located a three-brane, called the Standard Model (SM) brane where SM lives. 
The 5d fermions interact with the lepton and Higgs fields in the SM brane through a bulk-brane coupling parametrised by $\mu_1,\mu_2,\mu_3$. 
The left-handed neutrinos in SM acquire Dirac masses through the bulk-brane coupling. 
The model parameters are then
\begin{align}
R, \quad c_1, \quad c_2, \quad c_3, \quad \mu_1, \quad \mu_2, \quad \mu_3. 
\end{align}
We will demonstrate later on that we can actually determine $\mu_2$ and $\mu_3$ from the others by using the currrently available numerical values of the neutrino mass square differences.

Let us describe the setup concretely. The action consists of two parts:
\begin{align}
    S_5=S_{\mrm b}+S_{\mrm b\pd},
    \label{bulk_action}
\end{align}
where the 5D bulk action is
\begin{align}
    S_{\mrm b}=\int\!d^4x\int_0^{\pi R}\!dz\,\msum_{i=1}^3\Big(i\bar\Psi_i\Ga^M\pd_M\Psi_i-c_i\bar\Psi_i\Psi_i\Big),
    \label{bulk_quad}
\end{align}
and the bulk-brane interaction is
\begin{align}
    S_{\mrm b\pd}&=-M_*^{-1/2}\int\!d^4x\,\msum_{i=1}^3\Big(y_i\ol{\ell_i^{\text L}}\tilde H\Psi_i^\rR(z=0)+\text{c.c.}\Big),
\label{bulkbrane_quad}
\end{align}
where the bulk spacetime index $M$ runs over $0,1,2,3,4$ with $z=x^4$, $\pd/\pd z=\pd_4$ is understood, and $y_i$ are dimensionless coupling parameters, so that the dimensionful bulk-brane coupling is $y_i/\sqrt{M_*}$, with $M_*$ being the 5D Planck mass. 
$\ell_i^\rL$ stands for the SM lepton doublet $(\nu_i^\rL,e_i^\rL)$.
The label $i=1,2,3$ stands for the basis called the intermediate basis~\cite{Machado:2011jt,Carena:2017qhd}, 
in which both the bulk masses and the bulk-brane Yukawa couplings are diagonal.
This is different from the flavour basis
in which the weak interaction of the left-handed lepton doublets and the weak boson is diagonal, $\ol{\nu^\rL_\al}\ga^\mu e_\al^\rL W^+_\mu$ 
where $\al=\mrm{e},\upmu,\uptau$.
The left-handed neutrino in the flavour and intermediate bases are related by $\nu_\al^\rL=\sum_iU_{\al i}\nu_i^\rL$ 
with a $3\times3$ unitary matrix $U_{\al i}$ that plays a role of the Pontecorvo-Maki-Nakagawa-Sakata (PMNS) matrix.\footnote{
More details about the relation between the two bases are found in Appendix~\ref{app}.}
$\tilde H$ is the (conjugated) Higgs doublet which acquires the vacuum expectation value $(v,0)$ with $v=174$\,GeV upon the electroweak symmetry breaking.

As discussed in Appendix~\ref{app}, the bulk Dirac masses $c_i$ and the bulk-brane coupling constants $y_i$ come from diagonalisations of a
 Hermitian bulk mass matrix $\cC_{\al\bt}$ and a complex bulk-brane Yukawa matrix $\cY_{\al\bt}$ in the flavour basis, respectively. Therefore $c_i$ are real while $y_i$ are complex in general. The sign of $c_i$ cannot be absorbed and thus it is physical in contrast to 4D Dirac masses. 
On the other hand, by a similar argument to that on the Cabibbo-Kobayashi-Maskawa matrix in the SM quark sector, there remains one phase in the bulk-brane couplings, in contrast to the case with Majorana masses with three phases remaining.

Here we summarise our conventions about spinors.
The 5D Dirac matrices $\Ga^M$ are defined as $\Ga^\mu=\ga^\mu$ and $\Ga^4=-i\ga$, 
where $\mu$ runs over $0,1,2,3$. The 4D Dirac matrices $\ga^\mu$ satisfy $\{\ga^\mu,\ga^\nu\}=2\eta^{\mu\nu}I_{4\times 4}$ 
with $\eta_{\mu\nu}=\mrm{diag}\{1,-1,-1,-1\}$,
and $\ga$ is the 4D chirality matrix (we avoid the usual label $\gamma^5$ for clarity of indices)
such that $\{\ga,\ga\}=2I_{4\times 4}$, $\{\ga,\ga_\mu\}=0$ and $\ga^\dg=\ga$,
so that the 5D Dirac matrices satisfy $\{\Ga_M,\Ga_N\}=2\eta_{MN}I_{4\times 4}$ with $\eta_{4\mu}=0$ and $\eta_{44}=-1$.
As in 4D, it is convenient to decompose 5D spinors with the chirality matrix $\ga=i\Ga^4$ as
\begin{align}
    \Psi_i^\rL=\frac{1-\ga}{2}\Psi_i, \qquad \Psi_i^\rR=\frac{1+\ga}{2}\Psi_i,
\end{align}
so that $\ga\Psi_i^\rL=-\Psi_i^\rL$ and $\ga\Psi_i^\rR=\Psi_i^\rR$.

Boundary conditions on the 5D spinors at the two endpoints can be imposed only on half the spinor components since the left and right components are mixed. 
In our model, we impose Dirichlet boundary conditions that annihilate the left-handed part of $\Psi_i$:
\begin{align}
    \Psi_i^{\rL}(z=0)=\Psi_i^{\rL}(z=\pi R)=0.
    \label{Dbc}
\end{align}
To rewrite the action in terms of 4D fields, we prepare mode functions $f_{in}^{\rL}(z),f_{in}^{\rR}(z)$ for $\Psi_i^\rL,\Psi_i^\rR$ on the compact direction.
The 5D fields $\Psi_i^\rL,\Psi_i^\rR$ are then expanded as
\begin{align}
    \Psi_i^\rL(x,z)=\msum_n\psi_{in}^\rL(x)f_{in}^\rL(z), \qquad \Psi_i^\rR(x,z)=\msum_n\psi_{in}^\rR(x)f_{in}^\rR(z),
\end{align}
and the mode functions are required to be orthonormal:
\begin{align}
    &\int_0^{\pi R}dz f_{im}^a(z)f_{in}^b(z)^*=\de_{ab}\de_{mn}, \quad
    a,b\in\{\rL,\rR\}.
    \label{normalisation}
\end{align}
The boundary condition \eqref{Dbc} kills the left-handed zero mode $\psi_{i0}^\rL$ while the right-handed zero modes $\psi_{i0}^\rR$ survive. 
This is how chiral symmetry is broken in this model.
We then find the following mode functions: $f_{i0}^\rL=0$ by the boundary conditions and
\begin{align}
    f_{in}^\rL(z)&=\sqrt{\frac{2}{\pi R}}\sin\frac{nz}{R} \quad (n \geq 1), \\
    f_{i0}^\rR(z)&=\sqrt{\frac{2c_i}{e^{2\pi Rc_i}-1}}e^{c_iz}, \\
    f_{in}^{\rR}(z)&=\sqrt{\frac{2}{\pi R}}\frac{1}{\lm_{in}R}\left[c_iR\sin\left(\frac{nz}{R}\right)+n\cos\left(\frac{nz}{R}\right)\right] 
    \quad (n \geq 1),
\end{align}
where $\lm_{in}$ is given by
\begin{align}
    \lm_{in}=\sqrt{c_i^2+\frac{n^2}{R^2}},
\end{align}
and the overall constants are fixed by \eqref{normalisation}.
Equipped with the mode functions, we can rewrite the bulk action $S_{\mrm{b}}$ as a 4D action of the infinite KK towers of 4D spinors $\psi_{in}^\rL,\psi_{in}^\rR$:
\begin{align}
    \begin{split}        
S_{\mrm{b}\nu_{\rL}}
=\int\!d^4x\,\msum_{i=1}^3\Big[
&i\ol{\nu_i^{\rL}}\ol\sig{}^\mu\pd_\mu\nu_i^{\rL}+i\ol{\psi_{i0}^{\rR}}\sig^\mu\pd_\mu\psi_{i0}^{\rR}
+\msum_{n\geq 1}\left(i\ol{\psi_{in}^{\rL}}\ol\sig{}^\mu\pd_\mu\psi_{in}^{\rL}+i\ol{\psi_{in}^{\rR}}\sig^\mu\pd_\mu\psi_{in}^{\rR}\right) \\
&-\msum_{n\geq 1}\lm_{in}\left(\ol{\psi_{in}^{\rR}}\psi_{in}^{\rL}+\ol{\psi_{in}^{\rL}}\psi_{in}^{\rR}\right)\Big],
\end{split}
\label{4dKKaction-Sb}
\end{align}
where we included the kinetic term of the SM left-handed neutrinos $\nu_i^\rL$. 

In $S_{\mrm{b}\nu_{\rL}}$, the left-handed neutrinos $\nu_i^\rL$ replace the vanishing zero mode of the left-handed bulk fermions
and are still massless.
Their masses are generated by the bulk-brane interaction $S_{\mrm{b}\pd}$, 
which reads under the mode expansion:
\begin{align}
S_{\mrm{b}\pd}
&=-\int\!d^4x\,\msum_{i=1}^3\msum_{n=0}^\infty\left(Y_{in}\ol{\nu_i^{\text L}}\psi_{in}^{\text R}+Y_{in}^*\ol{\psi_{in}^{\text R}}\nu_i^{\text L}\right),
\label{bulkbrane_4D}
\end{align}
where $Y_{in}$ are defined by $Y_{in} = M_*^{-1/2}vy_if_{in}^{\text R}(0)$, reading concretely
\begin{align}
Y_{i0} = \mu_i \sqrt{\frac{2\pi Rc_i}{e^{2\pi R c_i}-1}}, \qquad
Y_{in} = \mu_i\sqrt{\frac{2 n^2}{n^2 + c_i^2 R^2}} \quad (n \geq 1),
\label{Y0i}
\end{align}
where we introduced the parameters $\mu_i$ of mass dimension $1$ as  
\begin{align}
\mu_i = M_*^{-\frac{1}{2}}\frac{v y_i}{\sqrt{\pi R}}. \label{mudef}
\end{align}
The bulk-brane interaction \eqref{bulkbrane_4D} adds a mass mixing between the SM left-handed neutrinos and the right-handed KK fermions to the diagonal mass terms in $S_{\mrm{b}\nu_\rL}$.
The physical masses are obtained by diagonalising the total mass matrix in \eqref{4dKKaction-Sb}$+$\eqref{bulkbrane_4D}, which is given explicitly in \eqref{mass-matrix}.
Here we summarise the result, relegating the details of the diagonalisation in Appendix~\ref{app}.
After diagonalisation, the total quadratic action reads:
\begin{align}
    \begin{split}
    S_{\mrm{b}\nu_\rL}+S_{\mrm{b}\pd}
    =\int\!d^4x\,\msum_{i=1}^3\msum_{n=0}^\infty&\left[
    i\ol{\nu_{i(n)}^\rL}\ga^\mu\pd_\mu \nu_{i(n)}^\rL+i\ol{\nu_{i(n)}^\rR}\ga^\mu\pd_\mu \nu_{i(n)}^\rR \right. \\
    &\qquad \left. 
    -m_{i(n)}\left(\ol{\nu_{i(n)}^\rL}\nu_{i(n)}^\rR+\ol{\nu_{i(n)}^\rR}\nu_{i(n)}^\rL\right)
    \right],
    \end{split} \label{nus-massbasis}
\end{align}
where $\nu^\rL_{i(n)},\nu^\rR_{i(n)}$ $(n=0,1,2,\cdots)$ denote the left- and right-handed parts of the 4D neutrinos and their KK excitations after diagonalisation, namely the 4D Dirac neutrinos in the mass basis. 
Their masses $m_{i(n)}$ are given as the positive roots of the equation
\begin{align}\label{transcendental}
        h_i(Rm_{i(n)})=0,
\end{align}
where the function $h_i(x)$ is given by
\begin{align}\label{hi-def}
h_i(x):=x^2 + \pi(R\mu_i)^2(Rc_i) - \pi(R\mu_i)^2\sqrt{x^2-(Rc_i)^2}\cot\Big(\pi\sqrt{x^2-(Rc_i)^2}\Big).
\end{align}
The masses $m_{i(n)}$ are labelled in the increasing order
\begin{align}
m_{i(0)}<m_{i(1)}<m_{i(2)}<\cdots.
\end{align}

As seen from the construction, only the SM left-handed neutrinos $\nu^\rL_i$ interact directly with the weak bosons. Therefore, the KK fermions $\psi^\rL_{in}$ $(n\geq1)$ are sterile.
The active left-handed neutrino in the flavour basis $\nu_\al^\rL$ in SM is related to the KK tower $\nu_{i(n)}^\rL$ in the mass basis by
\begin{align}
\nu_\al^\rL=\sum_{i=1}^3U_{\al i}\nu^\rL_i=\sum_{i=1}^3\sum_{n=0}^\infty U_{\al i}\cL^i_{0n}\nu_{i(n)}^\rL,
\end{align}
where the mixing matrix element $\cL^i_{0n}$ is given by
\begin{align}\label{mixmat-0a}
\cL^i_{0n}&=\left[
1+\frac{\bar c_i(\pi\bar\mu_i^2)^2+(Rm_{i(n)})^2\pi\bar\mu_i^2(1-2\pi\bar c_i-\pi^2\bar\mu_i^2)-\pi(Rm_{i(n)})^4}{2\pi\bar\mu_i^2[\bar c_i^2-(Rm_{i(n)})^2]}
\right]^{-\frac{1}{2}},
\end{align}
where we introduced the dimensionless version of the parameters of mass dimension $c_i,\mu_i$:
\begin{align}\label{dimless-cmu}
    \bar c_i:=c_iR, \qquad \bar \mu_i:=\mu_iR.
\end{align}
The other elements of the mixing matrix $\cL^i_{n\ell}$ are given by
\begin{align}\label{mixmat-na}
\cL^i_{n\ell}=\frac{\sqrt{2}\bar\mu_in}{(Rm_{i(\ell)})^2-\bar c_i^2-n^2}\cL^i_{0\ell} \quad (n \geq 1).
\end{align}
Note that ${\bar\mu}_i$ is related to the dimensionless bulk to brane coupling $y_i$ defined in \eqref{bulkbrane_quad} as
\begin{align}
\label{ymurel}
y_i=\sqrt{\pi}{\bar\mu}_i\sqrt{M_*R}/(vR)\simeq 10^{-2}\bar\mu_i(\text{eV}\,R)^{-1/2}\,,
\end{align}
where the numerical estimate corresponds to a typical value of the DD compactification scale around eV corresponding to $M_*\sim 10^9$ GeV.

\section{Mass spectrum}

The standard three-neutrino mixing model of neutrino masses has well described the oscillations of solar, atmospheric and reactor neutrinos.
Therefore, it is natural to consider extra dimensional effects in our model as a perturbation of the standard three-neutrino mixing model.
This observation motivates the requirement that the zero (i.e. lowest) mode masses of the three mass towers $m_{i(0)}$ 
are subject to the mass square differences that are determined through neutrino oscillations:
\begin{align}
    \De m_{ij}^2=m_{i(0)}^2-m_{j(0)}^2.
    \label{massdiff-zeromode}
\end{align}
Since neutrino oscillation experiments have constrained two types of mass square differences from solar and atmospheric neutrinos, 
they can reduce the model parameters by two via the relations \eqref{massdiff-zeromode}.

In the first two subsections, we present the structure of the solutions to the mass equations \eqref{transcendental} 
and explain how to generate the three mass towers, 
taking the mass square differences with \eqref{massdiff-zeromode} as physical data.
Then in the third subsection, we impose more physical information, in particular a larger compactification radius of $\cO(10)\,\mu$m and cosmological neutrino mass bounds, and derive some consequences.

\subsection{Structure of the mass equation}
\label{subsec:mass-structure}

\begin{figure*}
\begin{center}
  \includegraphics[width=1.0 \linewidth]{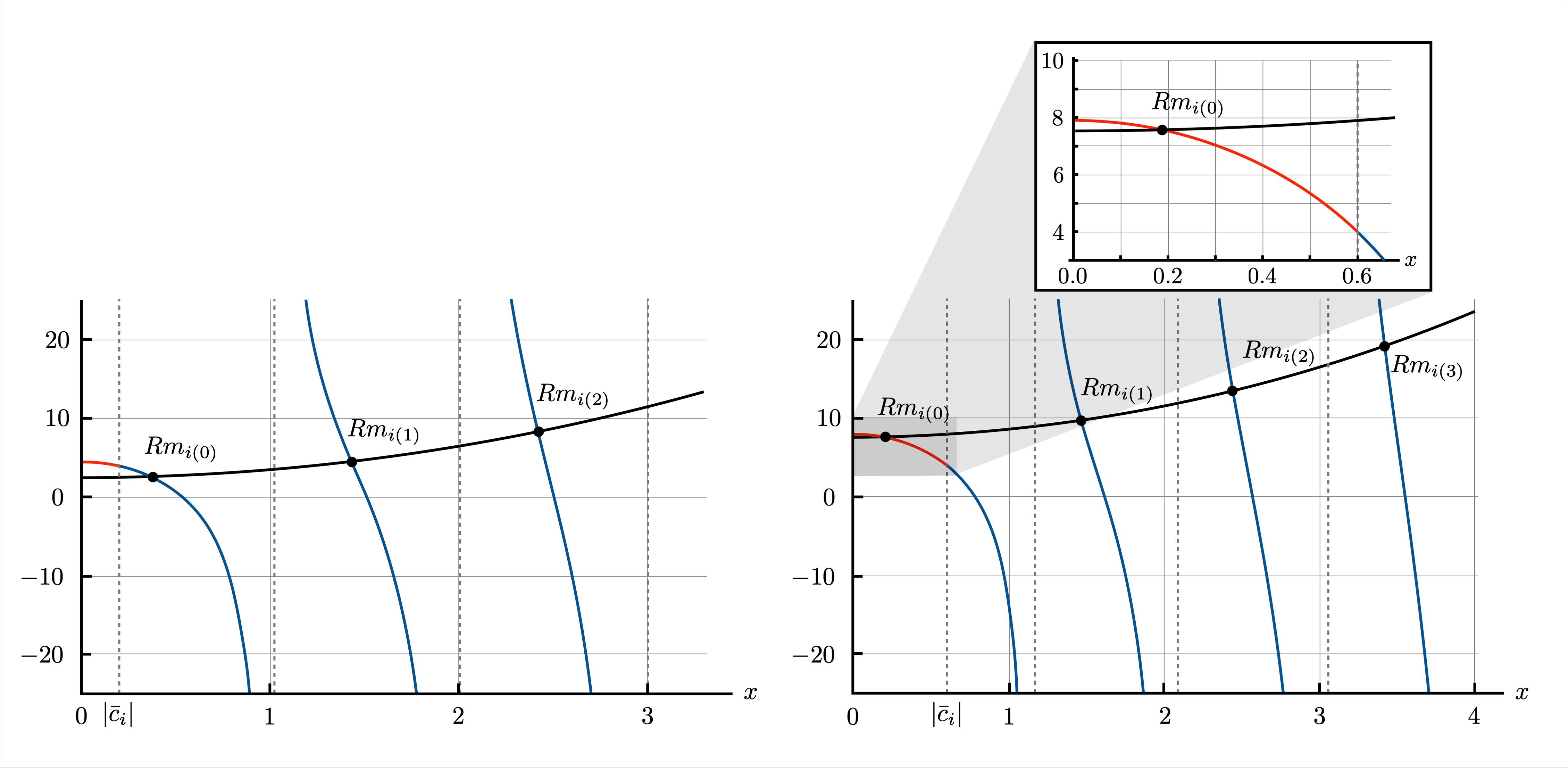}
 \caption{The two panels illustrate the roots of the mass equation \eqref{transcendental} as the intersections of $x^2+\pi\bar\mu_i^2\bar c_i$ (black curve) with the third term in \eqref{hi-def} as the red and blue curves for the $x<|\bar c_i|$ (coth) and $x>|\bar c_i|$ (cot) parts, respectively. The left panel shows the case $h(|\bar c_i|)<0$ where the coth part $x<|\bar c_i|$ has no solution, while the right panel shows the case $h(|\bar c_i|)>0$ where the coth part $x<|\bar c_i|$ has one solution.}
 \label{fig:trans_sols}
\end{center}
\end{figure*}

The KK masses $m_{i(n)}$ are the positive roots of the equation $h_i(Rm_{i(n)})=0$ \eqref{transcendental}.
The function $h_i(x)$ is defined not only for $x\geq|\bar c_i|$ as in \eqref{hi-def} but also for $x<|\bar c_i|$; For $x<|\bar c_i|$, the function $h_i$ reads \cite{Anchordoqui:2023wkm}
\begin{align}\label{hi-coth}
h_i(x)=x^2+\pi\bar\mu_i^2\bar c_i-\pi\bar\mu_i^2\sqrt{\bar c_i^2-x^2}\coth(\pi\sqrt{\bar c_i^2-x^2}).
\end{align}
We can therefore express the mass equation \eqref{transcendental} explicitly as follows:
\begin{alignat}{2}
&x^2+\pi\bar\mu_i^2\bar c_i-\pi\bar\mu_i^2\sqrt{x^2-\bar c_i^2}\cot(\pi\sqrt{x^2-\bar c_i^2})=0 &\quad \mbox{ for } ~ &x\geq|\bar c_i|, \label{transeq1}\\
&x^2+\pi\bar\mu_i^2\bar c_i-\pi\bar\mu_i^2\sqrt{\bar c_i^2-x^2}\coth(\pi\sqrt{\bar c_i^2-x^2})=0 &\quad  \mbox{ for } ~ &x<|\bar c_i|.  \label{transeq2} 
\end{alignat}
The solutions of $h_i(x)=0$ can then be classified into two cases: 
if $h_i(|\bar c_i|)>0$, it has one solution in the region $x<|\bar c_i|$ that solves the coth mass equation, 
while if $h_i(|\bar c_i|)<0$, it does not have a solution in this region, where $h_i(|\bar c_i|)$ is equal to
\begin{align}
h_i(|\bar c_i|)=\bar c_i^2+(\pi\bar c_i-1)\bar\mu_i^2.
\label{hi(cbari)}
\end{align}
This can be understood geometrically as in Figure~\ref{fig:trans_sols}, where the roots of the mass equation are the intersections of the parabola $x^2+\pi\bar c_i\bar\mu_i^2$ and the cot/coth curves (the third term of \eqref{hi-def}).
The parabola is monotonically increasing while the cot/coth curves consist of disjoint, monotonically decreasing curves.
When $h_i(|\bar c_i|)>0$, the parabola is above the coth/cot curves at $|\bar c_i|$ and intersects with both the cot and coth curves, 
while when $h_i(|\bar c_i|)<0$, the parabola is below the coth/cot curves at $|\bar c_i|$ and intersects only with the cot curves.

When $\pi\bar c_i>1$, the smallest root solves the coth mass equation because $h_i(|\bar c_i|)>0$ for any $\bar\mu_i$.
In contrast, when $\pi\bar c_i<1$, the criterion for the types of the mass equation solved by the smallest root is given by
\begin{align}
h_i(|\bar c_i|)>0 
~~\Longleftrightarrow~~ \bar\mu_i^2<\frac{\bar c_i^2}{1-\pi\bar c_i}
~~\Longleftrightarrow~~ \bar c_i \notin [\bar c_{i}^-, \bar c_{i}^+], 
\label{h>0} \\
h_i(|\bar c_i|)<0 
~~\Longleftrightarrow~~ \bar\mu_i^2>\frac{\bar c_i^2}{1-\pi\bar c_i}
~~\Longleftrightarrow~~ \bar c_i \in [\bar c_{i}^-, \bar c_{i}^+], 
\label{h<0}
\end{align}
where $\bar c_i^\pm$ are defined by
\begin{align}
\bar c_i^\pm = \frac{1}{2}\bar\mu_i \left(-\bar\mu_i\pi\pm\sqrt{\bar\mu_i^2\pi^2+4}\right).
\end{align}
Note that the condition $\pi\bar c_i<1$ still allows a negative $\bar c_i$ with large $|\bar c_i|$, which will be discussed in Sec.~\ref{subsec:condR}.
Below, we summarise the properties of the solutions in the two cases.

\subsubsection{Case $\bar c_i \in [\bar c_{i}^-, \bar c_{i}^+]$}
\label{subsub:cot}

Let us first look at the massless case $\bar c_i=0$. 
The roots solve the cot mass equation \eqref{transeq1} and live in the regions $n<Rm_{i(n)}<n+1/2$.
In general, the mass equation cannot be solved analytically.
However, when $\bar\mu_i^2$ is much larger or smaller than 1, the roots can be approximated by analytic expressions.
When $|\bar\mu_i|\ll1$, the roots are approximated by~\cite{Forero:2022skg}
\begin{align}
    m_{i(0)}\simeq\mu_i[1+\cO(\bar\mu_i^2)], \qquad
    m_{i(n)}\simeq\frac{n}{R}[1+\cO(\bar\mu_i^2)] \quad (n\geq1).
    \label{cot-small-mu}
\end{align}
We now turn to the opposite limit $\bar\mu_i\gg1$. Under this, the roots are approximated as\footnote{
    Here we computed up to the first order correction in $\bar\mu_i^{-2}$.}
\begin{align}
    m_{i(n)}=\frac{1}{R}\left(n+\frac{1}{2}\right)\left(1-\frac{1}{\pi^2\bar\mu_i^2}+\cO(\bar\mu_i^{-4})\right) \quad (n\geq0),
    \label{cot-large-mu}
\end{align}
which can be seen easily from Figure~\ref{fig:trans_sols}: the parabola becomes almost constant near zero.
Setting $\bar c_i=0$ in \eqref{mixmat-0a}, we find that the mixing matrix elements of the zero-mode (active) neutrino with its KK excitations $\cL^i_{0n}$ for large $|\bar\mu_i|\gg1$ read
\begin{align}
    \begin{split}
    &\cL^i_{0n}\simeq\frac{\sqrt{2}}{\pi\bar\mu_i} \quad(n\lesssim\bar\mu_i), \qquad
    \cL^i_{0n} \simeq \sqrt{\frac{2}{\pi}}\frac{\bar\mu_i}{n} \quad (n>\bar\mu_i).
    \end{split}
    \label{L0n-largemu}
\end{align}

Let us proceed to the massive case $c_i\neq0$. 
The roots solve the cot equation and live in the regions $\bar c_i^2+n^2<R^2m_{i(n)}<\bar c_i^2+(n+1)^2$,
which allows the following parametrisation:
\begin{align}
    R^2m_{i(n)}^2=\bar c_i^2+(n+y_n)^2, \quad 0<y_n<1.
\end{align}
$y_n$ decreases as $n$ grows because the parabola is monotonically increasing while each cot curve is monotonically decreasing.
The set of the roots $\{m_{i(n)}:n\geq0\}$ is classified into two types in terms of the region which the lowest root $m_{i(0)}$ belongs to: 
\begin{align}
    \text{type 1:} &\quad 0\leq y_0<\tfrac{1}{2} ~~\Longleftrightarrow~~ \pi\bar\mu_i^2\bar c_i+\bar c_i^2+\tfrac{1}{4}>0, \label{cot-type1} \\
    \text{type 2:} &\quad \tfrac{1}{2}\leq y_0<1 ~~\Longleftrightarrow~~ \pi\bar\mu_i^2\bar c_i+\bar c_i^2+\tfrac{1}{4}\leq0, \label{cot-type2}
\end{align}
where the condition for type 1 (type 2) arises when 
the parabola is above (below) zero when the leftmost cot curve crosses the zero at $x=\sqrt{\bar c_i^2+1/4}$ 
(see Figure~\ref{fig:trans_sols}). Type 2 is possible only when $c_i$ is negative.
We can then see that the other roots in each case satisfy
\begin{align}
    \text{type 1:} &\quad 0<y_n<\tfrac{1}{2} \quad \text{for all } n\geq1, \label{cot-all-type1} \\
    \text{type 2:} &\quad \tfrac{1}{2}\leq y_n<1 \quad \text{up to some } n_0\geq1; 
                    \quad 0<y_n<\tfrac{1}{2} \quad \text{for } n>n_0. \label{cot-all-type2}
\end{align}

We consider first the type 1 case \eqref{cot-all-type1}.
As in the $\bar c_i=0$ case, the mass equation cannot be solved analytically in general. 
However, if $y_n$ is small, we can expand the mass equation in $y_n$ and solve it for $y_n$ perturbatively. 
Let us then look for conditions on parameters for $y_n\ll1$.
For $n=0$, one can see from Figure~\ref{fig:trans_sols} that requiring $\vep^2:=-h_i(|\bar c_i|)\ll1$ yields $y_0\ll1$. 
Solving the mass equation perturbatively in $y_0$ yields $y_0=\vep^2$.
For $n\geq1$, note first that if $y_1\ll1$, then so are the other $y_n$ ($n\geq2$). Solving the mass equation perturbatively in $y_1$ yields
\begin{align}
    y_1=\frac{\bar c_i^2}{1-\pi\bar c_i}+\cO(\ep).
\end{align}
Requiring $y_1\ll1$ then yields $\pi|\bar c_i|\ll1$.
Moreover, since $y_n\ll1$ is guaranteed for all $n$, the roots can be written as
\begin{align}
    Rm_{i(n)}\simeq\sqrt{\bar c_i^2+n^2}+\cO(|\bar c_i|^2).
\end{align}
Note that $\vep^2\ll1$ is satisfied easily as long as $\pi|\bar c_i|\sim\pi\bar\mu_i\ll1$ is imposed.

In contrast, in type 2 case, the situation is different since $y_n>1/2$ up to some KK label according to \eqref{cot-all-type2}.
When $\bar\mu_i\gg1$, the parabola is almost constant $\sim\pi\bar\mu_i^2\bar c_i$ and the roots are the intersections of this line with the cot curves. 
In addition, if $|\bar c_i|\ll1$, then the intersection can be obtained perturbatively in $\bar c_i$.

\subsubsection{Case $\bar c_i \notin [\bar c_{i}^-, \bar c_{i}^+]$}
\label{subsub:coth}
When $\bar c_i \notin [\bar c_{i}^-, \bar c_{i}^+]$, each root is located in the following region:
\begin{align}\label{regions2}
&0<Rm_{i(0)}\leq|\bar c_i|, \\
&\sqrt{\bar c_i^2+n^2}<Rm_{i(n)}<\sqrt{\bar c_i^2+\left(n+\tfrac{1}{2}\right)^2} \quad ~~ (n \geq 1).
\end{align}
The smallest root $Rm_{i(0)}$ of the mass equation is smaller than $|\bar c_i|$ and hence solves the coth equation \eqref{transeq2}, 
while the others are greater than $|\bar c_i|$ and solve the cot equation \eqref{transeq1}.
As can be seen from Figure~\ref{fig:trans_sols}, when $\bar c_i$ gets larger, the parabola goes up and the intersection with the coth curve gets smaller, 
while the other roots are still greater than $\bar c_i$  as they are all on the cot curve. 

Let us quantify this argument. 
The requirement $Rm_{i(0)}\ll|\bar c_i|$ is equivalent to
\begin{align}\label{h0<<hc}
    |h_i(0)|\ll|h_i(|\bar c_i|)|.
\end{align} 
Under this, we can solve \eqref{transeq2} after expanding it in $x^2$:
\begin{align}\label{approx-cothroot-0}
    m_{i(0)}^2\simeq\frac{2\pi\mu_i^2|\bar c_i|(|\bar c_i|\coth(\pi|\bar c_i|)-\bar c_i)}
    {2|\bar c_i|+\pi\bar\mu_i^2\coth(\pi|\bar c_i|)-\pi^2\bar\mu_i^2|\bar c_i|/\sinh^2(\pi|\bar c_i|)}.
\end{align}
We can simplify this further by imposing the following condition:
\begin{align}
    \begin{split}
        |\bar c_i|\gg\max\{1, \, \bar\mu_i, \, \pi\bar\mu_i^2\},
    \end{split}
    \label{approx-cond}
\end{align}
which guarantees \eqref{h0<<hc}.
Then, the approximate solution \eqref{approx-cothroot-0} is simplified into
\begin{align}
    m_{i(0)}\simeq
    \begin{cases}
        \mu_i\sqrt{2\pi|\bar c_i|}e^{-\pi\bar c_i} & \text{when } c_i>0, \\
        \mu_i\sqrt{2\pi|\bar c_i|} & \text{when } c_i<0.
    \end{cases}
    \label{approx-cothroot}
\end{align}
The condition \eqref{approx-cond} indeed guarantees $Rm_{i(0)}\ll|\bar c_i|$.

Let us next consider the higher KK modes $m_{i(n)}$ for $n\geq1$ that solve the cot mass equation \eqref{transeq1}.
Its roots $Rm_{i(n)}$ satisfy $R^2m_{i(n)}^2=\bar c_i^2+(n+y_n)^2$ with $0<y_n<1/2$.
The condition $h_i(|\bar c_i|)\gg0$ yields $0<y_n\ll1$, and $y_n$ gets smaller as $n$ increases.
This property allows the expansion of the cot equation \eqref{transeq1} up to first order in $y_n^2$. 
Solving it for $y_n$ gives
\begin{align}\label{al_i(n)}
    Rm_{i(n)} \simeq \sqrt{\bar c_i^2+n^2}\left[1+\frac{n^2\bar\mu_i^2}{(\bar c_i^2+n^2)^2}+\cdots\right].
\end{align}
Under the condition \eqref{approx-cond}, the second term in the bracket can be neglected for all $n$.
Note that \eqref{approx-cond} yields \eqref{h0<<hc}, which satisfies the condition $h_i(|\bar c_i|)\gg0$ used here.

Under the condition \eqref{approx-cond}, the mixing matrices $\cL^i_{0n}$ in \eqref{mixmat-0a} are also simplified.
For higher KK modes $n\geq1$, by using $R^2m_{i(n)}^2\simeq\bar c_i^2+n^2$ from \eqref{al_i(n)}, we obtain
\begin{align}
    \cL^i_{0n}\simeq\frac{\sqrt{2}\bar\mu_in}{\bar c_i^2+n^2}.
    \label{Li0n}
\end{align}
The maximum of $\cL^i_{0n}$ is ${\bar\mu_i}/(\sqrt{2}|\bar c_i|)$ at $n\simeq|\bar c_i|$.
Since $|\bar c_i|$ is large, we can approximate the sum of $|\cL^i_{0n}|^2$ over $n\geq1$ 
by an integral under the replacement $n/|\bar c_i|\to x$:
\begin{align}
    \msum_{n=1}^\infty|\cL^i_{0n}|^2
    \simeq \frac{2\bar\mu_i^2}{|\bar c_i|}\int_0^\infty\frac{x^2}{(1+x^2)^2}dx
    =\frac{\pi\bar\mu_i^2}{2|\bar c_i|}.
\end{align}
The unitarity relation $1=\sum_{n=0}^\infty|\cL^i_{0n}|^2$ gives 
\begin{align}\label{Li00-approx}
    |\cL^i_{00}|\simeq1-\frac{\pi\bar\mu_i^2}{4|\bar c_i|}.
\end{align}
Note that the sum of $|\cL^i_{0n}|^2$ over higher KK modes is suppressed under the condition \eqref{approx-cond}, 
which can suppress large deviation from the results of neutrino oscillation experiments due to the active-to-sterile disappearance effect
and justify our requirement for the zero mode masses and the mass square differences.

\subsection{Constraints by mass square differences}

As already explained, we require that the zero mode masses are constrained by the mass square differences from the oscillation of the solar and atmospheric neutrinos.
This implies that once the mass square differences and one of the three zero modes, say $Rm_{1(0)}$, are given, the other two zero modes are fixed.
Let us suppose that we have solved the mass equation $h_1(x)=0$ with parameters $(R,c_1,\mu_1)$ and obtained the roots $m_{1(n)}$. 
Once the mass square differences $\De m_{i1}^2$ are given, the zero mode masses for $i=2,3$ are fixed to
\begin{align}
    m_{i(0)}^2=m_{1(0)}^2+\De m_{i1}^2. \label{mi2_Dem2}
\end{align}
According to the global fit~\cite{Esteban:2024eli}, the central values of the mass square differences are given by
\begin{align}
    \text{normal hierarchy (NH)}: &~~ \De m_{21}^2=7.49\times10^{-5}\,\text{eV}^2, ~~ \De m_{31}^2=2.534\times10^{-3}\,\text{eV}^2, 
    \label{massdiff-NH} \\
    \text{inverted hierarchy (IH)}: &~~ \De m_{21}^2=7.49\times10^{-5}\,\text{eV}^2, ~~ \De m_{32}^2=-2.510\times10^{-3}\,\text{eV}^2,
    \label{massdiff-IH}
\end{align}
where in IH $\De m_{31}^2=\De m_{21}^2+\De m_{32}^2$.
On the other hand, the two zero modes $Rm_{2(0)},Rm_{3(0)}$ are defined as the smallest roots of the mass equations
$h_i(Rm_{i(0)})=0$ for $i=2,3$, which fixes $\bar\mu_i^2$. Concretely, for each $i=2,3$:
\begin{itemize}
    \item If $Rm_{i(0)}\leq|\bar c_i|$, the condition $h_i(Rm_{i(0)})=0$ gives
    \begin{align}
        \bar\mu_i^2=\frac{R^2m_{i(0)}^2}{\pi\sqrt{\bar c_i^2-R^2m_{i(0)}^2}\coth\Big(\pi\sqrt{\bar c_i^2-R^2m_{i(0)}^2}\Big)-\pi\bar c_i}.
        \label{mui-coth}
    \end{align}
    Since $\bar\mu_i^2>0$, we obtain the following consistency condition:
    \begin{align}
        \sqrt{\bar c_i^2-R^2m_{i(0)}^2}\coth\Big(\pi\sqrt{\bar c_i^2-R^2m_{i(0)}^2}\Big)>\bar c_i.
        \label{mui-consis-coth}
    \end{align}

    \item If $Rm_{i(0)}>|\bar c_i|$, the condition $h_i(Rm_{i(0)})=0$ gives
    \begin{align}
        \bar\mu_i^2=\frac{R^2m_{i(0)}^2}{\pi\sqrt{R^2m_{i(0)}^2-\bar c_i^2}\cot\Big(\pi\sqrt{R^2m_{i(0)}^2-\bar c_i^2}\Big)-\pi\bar c_i}.
        \label{mui-cot}
    \end{align}
    In this case, the consistency conditions are not only the positivity $\bar\mu_i^2>0$ but also the requirement that $Rm_{i(0)}$ is the smallest root:
    \begin{align}
        \begin{split}
        &\sqrt{R^2m_{i(0)}^2-\bar c_i^2}\cot\Big(\pi\sqrt{R^2m_{i(0)}^2-\bar c_i^2}\Big)>\bar c_i, \\
        &\sqrt{\bar c_i^2+\bt^2}\leq Rm_{i(0)}<\sqrt{\bar c_i^2+(\bt+\tfrac{1}{2})^2}
        \end{split}
        \label{mui-consis-cot}
    \end{align}
    where $\bt=0$ ($1/2$) when $m_{i(0)}$ is of type 1 (2).
\end{itemize}
Once $\bar\mu_2,\bar\mu_3$ are obtained, we can solve the equations $h_2(x)=h_3(x)=0$ to obtain the rest of the roots $Rm_{2(n)},Rm_{3(n)}$ $(n\geq 1)$, 
which solve the cot mass equations \eqref{transeq1}. 

In summary, once the parameters $R,c_1,c_2,c_3$ and one of $\bar\mu_i$ and the two mass square differences are given, 
we can obtain all KK masses $m_{i(n)}^2$. In NH, we choose $\bar\mu_1$ as a given parameter and determine $\bar\mu_2,\bar\mu_3$
since the parameter $\bar\mu_1$ corresponds to the lightest zero mode mass~\cite{Forero:2022skg}.
On the other hand, it is natural in IH to choose $\bar\mu_3$ as a given parameter and determine $\bar\mu_1,\bar\mu_2$
through $m_1^2=m_3^2-\De m_{31}^2$ and $m_2^2=m_1^2+\De m_{21}^2$~\cite{Forero:2022skg}. 
The relations and conditions \eqref{mui-coth}--\eqref{mui-consis-cot} hold again as they are except that the index $i$ refers to $i=1,2$.

\subsection{General consequences from physical inputs}
\label{subsec:condR}

Based on mathematical results derived in the previous subsections, we proceed to obtain consequences by imposing physical constraints. One is the possibility of having compactification radius of order $R\sim\cO(10)\,\mu$m within the range of the Dark Dimension proposal, below the upper bound from table top experiments~\cite{Lee:2020zjt},
\begin{align}
    R\lesssim30\,\mu\text{m}\, .
    \label{R-ub}
\end{align}
Another one is an upper bound on the neutrino masses which comes from cosmology, according to Planck 2018 results~\cite{Planck:2018nkj,Planck:2018vyg},
\begin{align}
    m_{1(0)}+m_{2(0)}+m_{3(0)}<0.12\,\text{eV}.
    \label{Planck-ub}
\end{align}
A goal here is to obtain as implication that requiring $R\sim\cO(10)\,\mu$m favours larger bulk masses $|\bar c_i|\gg1$.
For this, we are deriving several consequences on the model parameters by using \eqref{R-ub} and \eqref{Planck-ub} together with the mass squared differences from neutrino oscillations \eqref{massdiff-NH}, \eqref{massdiff-IH}.

\subsubsection{Bounds on the compactification radius: cot case}
\label{subsubsec:ubR-cot}

We first derive upper bounds on the compactification radius and then obtain their implications.
Let us first suppose that the zero mode masses satisfy the cot mass equations \eqref{transeq1}.
In the following, we restrict for concreteness to the NH case, while the IH can be analysed in a similar way.
\paragraph{General bounds.}
We first derive some general upper bounds.
For this, we proceed as follows: 
\begin{align}
    \begin{split}
    &|\bar c_1|^2+\bt_1^2+R^2\De m_{i1}^2
    <R^2m_{1(0)}^2+R^2\De m_{i1}^2
    =R^2m_{i(0)}^2
    <|\bar c_i|^2+\bt_i^2 \\
    ~~\Longrightarrow~~ &
    R^2\De m_{i1}^2<|\bar c_i|^2-|\bar c_1|^2+\bt_i^2-\bt_1^2,
    \label{inequality-cot}
    \end{split}
\end{align}
where $\bt_1=0$ or $1/2$ (for type 1 or 2 $m_{1(0)}$), 
and $\bt_i=1/2$ or 1 (for type 1 or 2 $m_{i(0)}$) for $i=2,3$,
according to the definitions \eqref{cot-type1} and \eqref{cot-type2}.
Note that this leads to $|\bar c_i|^2-|\bar c_1|^2+\bt_i^2-\bt_1^2>0$.

Let us derive concrete upper bounds. Before proceeding, recall that $\pi\bar c_i<1$.
\begin{itemize}
\item When $c_2=c_3=0$, the mass equations for $i=2,3$ are both the cot one and hence we have 
\begin{align}
    R^2\De m_{i1}^2<R^2m_{1(0)}^2+R^2\De m_{i1}^2=R^2m_{i(0)}^2<\frac{1}{4}
    ~~\Longrightarrow~~
    R<\frac{1}{2\sqrt{\De m_{31}^2}},
    \label{c=0-ubR}
\end{align}
due to $\De m_{21}^2<\De m_{31}^2$.
Using the concrete values in \eqref{massdiff-NH}, we obtain $R<2\,\mu$m~\cite{Forero:2022skg}.
\item Let us move on to the massive case $c_i\neq0$. 
We first consider the case $-1<\pi\bar c_i<1$,
where we can derive a model independent upper bound: since $\bt_i^2-\bt_1^2\leq1$, we obtain
\begin{align}
    R^2\De m_{i1}^2<|\bar c_i|^2-|\bar c_1|^2+1
    <\frac{1}{\pi^2}+1 ~~\Longrightarrow~~
    R<\sqrt{\frac{1}{\De m_{i1}^2}\left(\frac{1}{\pi^2}+1\right)}.
\end{align}
Using the concrete values in \eqref{massdiff-NH}, we obtain
\begin{align}
    i=2: &\quad R<23.9\,\mu\text{m}, \label{cot2-Rubd} \\
    i=3: &\quad R<4.1\,\mu\text{m}. \label{cot3-Rubd}
\end{align} 

\item In contrast, when $\pi\bar c_i<-1$, we can obtain another upper bound on $|\bar c_i|$ from the cosmological bound \eqref{Planck-ub}: Since $Rm_{i(0)}>|\bar c_i|$ for the cot mass equation, we obtain
$-0.12R<\bar c_i$, which, however, does not give an upper bound.\footnote{In a similar manner, one might derive 
\begin{align}
    R<\sqrt{\frac{1}{\De m_{i1}^2}\left(0.12^2R^2+\bt_i^2-\bt_1^2\right)},
\end{align}
but it is satisfied for any $R$.}
We may therefore just use the general inequality \eqref{inequality-cot} as it is. 
The upper bound on $R$ then depends on concrete choices of the bulk masses. 
For example, when $|\bar c_1|=|\bar c_i|$, we obtain (again using $\bt_i^2-\bt_1^2\leq1$)
\begin{align}
    i=2: &\quad R<22.8\,\mu\text{m}, \label{cot2-Rubd-pic<-1} \\
    i=3: &\quad R<3.9\,\mu\text{m}. \label{cot3-Rubd-pic<1}
\end{align} 
\end{itemize}

\paragraph{Summary and implications: cot case.}
Let us summarise the arguments above with some implications on our model parameters.
\begin{itemize}
    \item If $\pi|\bar c_i|<1$ ($i=2,3$) and $m_{i(0)}$ solves the cot mass equation, then the compactification radius is upper bounded as in \eqref{cot2-Rubd} and \eqref{cot3-Rubd}. In particular, if $\pi|\bar c_i|\ll1$, the condition $h_i(|\bar c_i|)<0$ is satisfied for almost all $\bar\mu_i$ according to the second condition in \eqref{h<0}, which means that $m_{i(0)}$ solves the cot mass equation automatically. These results imply that if we want to explore the compactification radius around $\cO(10)\,$eV as indicated by Dark Dimension scenario, larger $|\bar c_i|>1$ are favoured.

    \item On the other hand, if $\pi\bar c_i<-1$ and $m_{i(0)}$ solves the cot mass equation, then the upper bound on $R$ depends on the concrete bulk masses.
    Let us consider the case  where the bulk masses are equal up to signs $|c|=|c_i|$, which will be discussed in the next section. 
    Then the upper bound on $R$ is \eqref{cot2-Rubd-pic<-1} and \eqref{cot3-Rubd-pic<1}. 
    In this case, for larger radius around $\cO(10)\,\mu$m, the mass $m_{3(0)}$ may have to solve the coth mass equation instead, which will be discussed again at the end of this subsection.
    On top of this, the cosmological bound can constrain $\bar c$. 
    Since $Rm_{1(0)}>|\bar c|$ irrespective of the type of $m_{1(0)}$, the Planck bound \eqref{Planck-ub} implies $R>1.6{|\bar c|}\,\mu\text{m}$.
    On the other hand, $m_{3(0)}$ satisfies (with $\bt_1=0$ and $\bt_3=1$ in \eqref{inequality-cot} for a universal bound)
    \begin{align}
        |\bar c|^2+R^2\De m_{31}^2<R^2(m_{1(0)}^2+\De m_{31}^2)=R^2m_{3(0)}^2<|\bar c|^2+1,
    \end{align}
    which gives $R<3.9\,\mu$m. Combining this with the lower bound $R>1.6{|\bar c|}\,\mu\text{m}$ just obtained leads to $|\bar c|<2.4$.
    This indicates that  for larger $|\bar c|$, cases with zero modes solving the coth mass equations will be preferable.

\end{itemize}

\paragraph{Case of massless bulk fermions with large $\boldsymbol{\bar\mu_i}$.}
\label{subsubsec:largemu-Rub}
    Let us move on to a quite different case: $\bar\mu_i\gg1$ with massless bulk fermions $c_1=0$, which was discussed partly in Sec.~\ref{subsub:cot}. 
    In this case, the masses are given by \eqref{cot-large-mu}. 
    The cosmological bound \eqref{Planck-ub} implies $m_{i(0)}<0.12$\,eV, which gives a lower bound on $R>0.82\,\mu$m (since $m_{i(0)}\approx 1/(2R)$).
    Combining this with the upper bound $R<2\,\mu$m derived from the mass square differences in Sec.~\ref{subsubsec:ubR-cot}, it implies 
    \begin{align}\label{lb-ub-R-largemu}
        0.82\,\mu\text{m}<R<2.0\,\mu\text{m}.
    \end{align}
    
    A previous work for $|\bar c_i|=0$ found an upper bound $R\simlt 0.4\,\mu$m by combining results of neutrino oscillation experiments, under the assumption of small $\bar\mu_i^2$~\cite{Forero:2022skg}. 
    However, as $\bar\mu_i$ gets larger, the mixing of the zero-mode active neutrino with the higher KK modes gets suppressed as \eqref{L0n-largemu}. 
    Then the extra-dimensional effects become negligible, so the upper bound $R\simlt 0.4\,\mu$m is not reliable and opens a new allowed narrow window of larger $R$ satisfying the bound we found in \eqref{lb-ub-R-largemu}.

\subsubsection{Bounds on the compactification radius: coth case}
\label{subsubsec:ubR-coth}

\paragraph{General bounds.}
Next, we investigate the case where $m_{i(0)}$ with $i=2$ or 3 satisfies the coth mass equation \eqref{transeq2}.
We can also derive an upper bound on $R$ in this case, 
but for this, we need to analyse the positivity of $\bar\mu_i^2$ given in \eqref{mui-consis-coth} in detail.
It is convenient to introduce the following function: 
\begin{align}
    d_i(\al):=\sqrt{\bar c_i^2-\al^2}\coth\Big(\pi\sqrt{\bar c_i^2-\al^2}\Big)-\bar c_i.
\end{align}
which is defined for $0\leq\al\leq|\bar c_i|$, and is monotonically decreasing for fixed $\bar c_i$. Note that $d(\bar c_i)=\pi^{-1}-\bar c_i$.
When $\bar c_i<\pi^{-1}$, $d_i(\al)$ is positive and hence the positivity of $\bar\mu_i^2$ \eqref{mui-consis-coth} is satisfied.

On the other hand, when $\bar c_i>\pi^{-1}$, since $d_i(0)>0$ but $d_i(\bar c_i)<0$,
$d_i(\al)$ has only one root, which we denote by $\bar\al(\bar c_i)$, and $d_i(\al)$ is positive for $\al<\bar\al(\bar c_i)$.
Therefore, $\bar\mu_i^2$ is positive if and only if
\begin{align}\label{ali0<baral}
    Rm_{i(0)}<\bar\al(\bar c_i).
\end{align}
When $\bar c_i$ increases, $\bar\al(\bar c_i)$ monotonically decreases exponentially to zero. 
Since $\bar c_i>\pi^{-1}$ and $\bar\al(\pi^{-1})=\pi^{-1}$, we have 
\begin{align}
    0<\bar\al(\bar c_i)<\pi^{-1}.
\end{align}
Now, assuming that $\bar c_i>\pi^{-1}$, we can derive an upper bound on $R$.
Since $Rm_{i(0)}$ solves the coth mass equation, 
the positivity of $\bar\mu_i^2$ dictates that \eqref{ali0<baral} should hold:
\begin{align}
    R^2m_{i(0)}^2<\bar\al(\bar c_i)^2.
\end{align}
Since $\De m_{i1}^2<m_{i(0)}^2$ and $\bar\al(\bar c_i)<\pi^{-1}$, we obtain
\begin{align}
    R^2\De m_{i1}^2<\pi^{-2}.
\end{align}
Therefore, for $i=2$ or 3, if $\bar c_i>1/\pi$ and $Rm_{i(0)}$ solves the coth mass equation, $R$ is bounded from above as:
\begin{align}
    i=2: &\quad R<7.25\,\mu\text{m}, \label{coth2-Rubd} \\
    i=3: &\quad R<1.25\,\mu\text{m}. \label{coth3-Rubd}
\end{align}

\paragraph{Summary and implications: coth case.}
Let us apply the above results to a concrete case
$|c|:=|c_1|=|c_2|=|c_3|\neq0$, which will be discussed in the next section.
Since $m_{2(0)}<m_{3(0)}$, we have three patterns:
\begin{align}
    (m_{2(0)},m_{3(0)}) &: \quad (\cot, \cot), \nn\\
    (m_{2(0)},m_{3(0)}) &: \quad (\coth, \cot), \nn\\
    (m_{2(0)},m_{3(0)}) &: \quad (\coth, \coth). \nn
\end{align}
In the first two cases, $m_{3(0)}$ solves the cot mass equation 
and thus the compactification radius is forced to be smaller than $4.1\,\mu$m due to \eqref{cot3-Rubd} or $3.9\,\mu$m due to \eqref{cot3-Rubd-pic<1}.
Therefore, if we are interested in larger radius, both $m_{2(0)}$ and $m_{3(0)}$ must solve the coth mass equations.
In particular, for $R>3.9\,\mu$m, we need $\bar c<\pi^{-1}$ due to \eqref{coth3-Rubd}.
However, if $|\bar c|$ is so small that $|\bar c|<\pi^{-1}$, the requirement $R>3.9\,\mu$m implies $|c|<0.016$\,eV.
On the other hand, in the next section, we will need to explore much larger $|c|$. In conclusion, if we want to explore larger $|c|$ with the requirement of larger radius, we have to choose negative bulk masses such that $\bar c<-\pi^{-1}$.

\section{Differential beta decay spectrum}

The Karlsruhe Tritium Neutrino (KATRIN) experiment~\cite{KATRIN:2021dfa} aims at determining the neutrino mass scale 
by analyzing the $\beta$-decay of molecular tritium ${\rm T}_2 \rightarrow {}^3{\rm HeT}^+ + e^- +\bar\nu_e$.
The combination of a high-luminosity windowless gaseous molecular tritium source with a high-resolution spectrometer enables KATRIN 
to perform a precise measurement of the energy spectrum of electrons from the beta decay down to $40~\text{eV}$ 
below the endpoint of the spectrum at $E_0 \approx 18.57 ~\text{keV}$
where the impact of the neutrino mass is maximal. 
Recently, KATRIN has published the upper limit on the effective neutrino mass at $0.45~\text{eV}$  ($90 \%$ C.L.) \cite{KATRIN:2024cdt}.

If the electron neutrino is a linear combination of three eigenstates of mass eigenvalues $m_i$ $(i=1,2,3)$ with the PMNS matrix components $U_{\mrm{e}i}$, 
the differential spectrum for $\beta$ electrons with kinetic energy $E$ is expressed by\footnote{
   For the $\beta$-decay of a molecular tritium, the effect of final states of the daughter molecular system should be added.
   Modifications by this, however, do not affect the discussion in what follows. 
   More details about such effects and other corrections are found in, for example,~\cite{Kleesiek:2018mel}.}  
\begin{align}
    \cR_{\mrm{std}}&=\cC\sum_{i=1}^3|U_{\mrm{e}i}|^2\ep\sqrt{\ep^2-m_i^2}\;\te(\ep-m_{i}),
    \label{std_decay_rate}
\end{align}
where $\cC$ is a common factor independent of models of neutrino masses, 
$\ep=E_0-E$, and $\te(x)$ is the step function defined by $\te(x)=1$ for $x>0$ and $\te(x)=0$ for $x<0$.
Note that the mass observed in KATRIN is not the mass of an active neutrino eigenstate $m_i$, 
but an effective neutrino mass defined by
\begin{align}    
    m_\nu:=\left(\sum_{i=1}^3|U_{\mrm{e}i}|^2m_{i}^2\right)^{1/2}.
    \label{m_nu_eff}
\end{align}
This definition is motivated by the fact that 
the difference of the active neutrino eigenstates is below the sensitivity of the electron energy (lower bound of $\ep$) $0.4$\,eV
and thus negligible.
Concretely, expanding $\sqrt{\ep^2-m_{i}^2}$ in $m_{i}^2/\ep^2$, we can rewrite
\begin{align}
    \cR_{\mrm{std}}\simeq\cC\sum_{i=1}^3|U_{\mrm{e}i}|^2(\ep^2-m_{i}^2/2)\te(\ep-m_{i}).
\end{align}
Since the difference of $m_\nu$ and $m_i$ is smaller than the energy sensitivity, we can replace $m_i$ in the step function by $m_\nu$.
The result is
\begin{align}
    \cR_{\mrm{std}}
    \simeq\cC(\ep^2-m_\nu^2/2)\te(\ep-m_\nu)
    \simeq\cC\ep\sqrt{\ep^2-m_\nu^2}\,\te(\ep-m_\nu).
\end{align}
KATRIN obtains $m_\nu$ as the neutrino mass through this equation.

If a hypothetical sterile eigenstate of mass $m_4$ contributes to the electron neutrino on top of the three mass eigenstates,
then the sterile eigenstate can be produced with a probability determined by a mixing between electron neutrino and sterile neutrino state. 
For example, in the case of a sterile neutrino in the 3+1 model, the differential spectrum becomes
\begin{align}
    \cR_{3+1}=(\cC\cos^2\theta_{\mrm{ee}})~\ep\sqrt{\ep^2-m_\nu^2}\,\te(\ep-m_\nu)
    +(\cC\sin^2\theta_{\mrm{ee}})~\ep\sqrt{\ep^2-m_4^2}\,\te(\ep-m_4),
    \label{3+1_decay_rate}
\end{align}
where $\sin^2(\theta_{\mrm{ee}})$ is the mixing angle between active and sterile states. 
In terms of observation, a sterile neutrino would leave two important imprints on the electron spectrum: a global distortion and a clear kink signal. 
The mixing of sterile and active neutrino determines the amplitude of the global distortion. 
The position of the kink allows determination of the fourth neutrino mass as it occurs at an energy equal to $E_0-m_4$.

In the case of our model with Dark Dimension, the electron neutrino is a linear combination of infinitely many KK states 
on top of the three zero modes. 
If these KK neutrinos are emitted during the $\beta$-decay, it is natural to expect more kinks to occur at energies $E_0-m_{i(n)}$ for $n\geq 1$. 
We explore this possibility in what follows.

\subsection{Differential beta decay spectrum and continuous approximation of KK spectrum}

\begin{figure*}
\begin{center}
  \includegraphics[width=0.70 \linewidth]{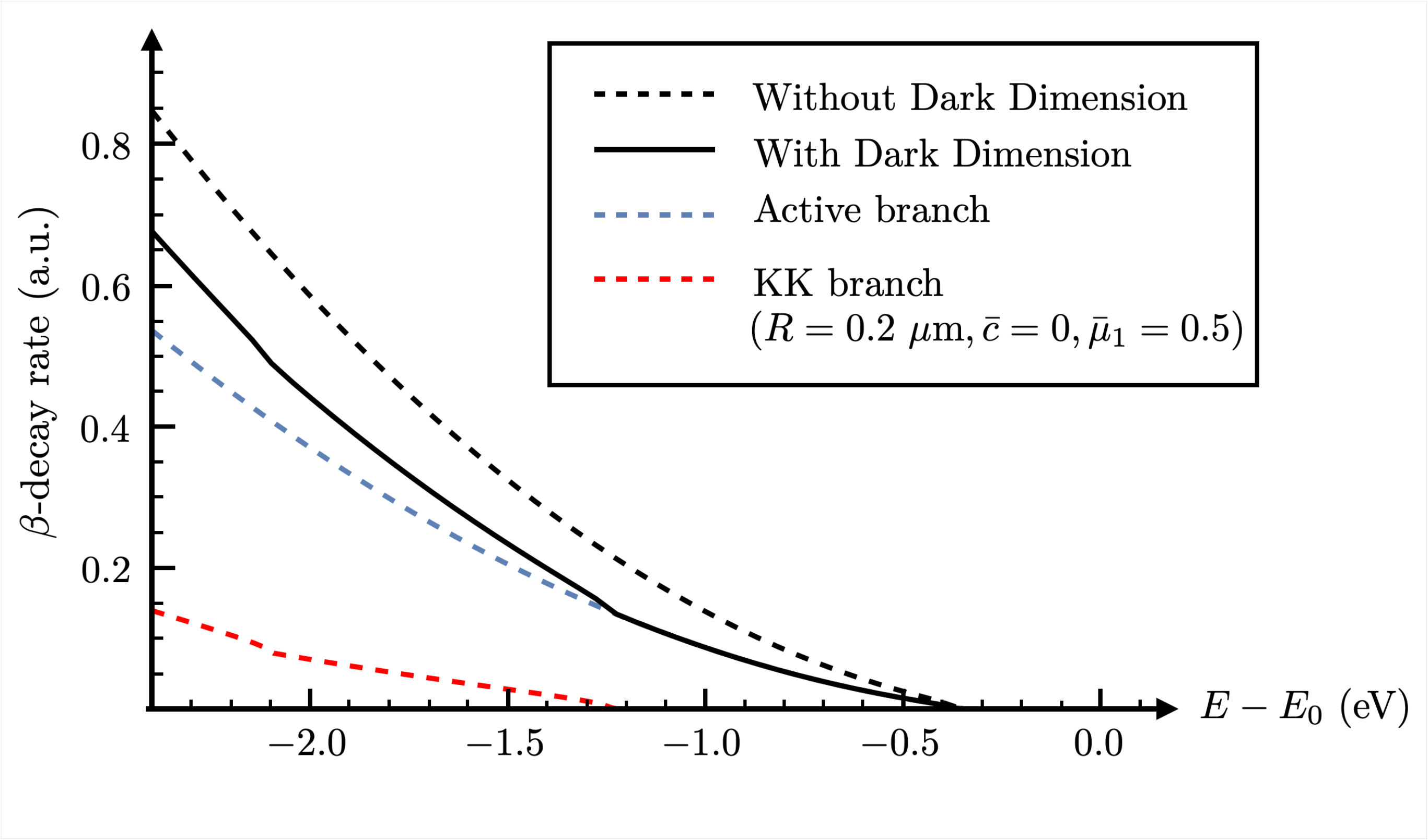}
 \caption{For an extra dimension with radius $R=0.2~\mu\text{m}$, the region near the endpoint of the spectrum is distorted due to the KK modes. In this  case, KK modes with $n=1,2$ contribute. We can see the two characteristic ``kink" signatures on the spectrum.}
 \label{sub_micron_KK}
\end{center}
\end{figure*}

In our model with bulk fermions, the electron neutrino mixes with the mass eigenstates as 
$\nu_\mrm{e}^\rL=\sum_{i=1}^3\sum_{n\geq0}U_{\mrm{e}i}\cL^i_{0n}\nu^\rL_{i(n)}$.
Therefore, the differential spectrum is modified into
\begin{align}
    \cR_\bt(\ep,\{m_{i(n)}\})=\cR_0+\cR_{\text{KK}}, 
\end{align}
where we separated the contribution of the zero modes from the sum over KK label $n$:
\begin{align}
    \cR_0&=\cC\sum_{i=1}^3|U_{\mrm{e}i}\cL^i_{00}|^2\ep\sqrt{\ep^2-m_{i(0)}^2}\te(\ep-m_{i(0)}), \\
    \cR_{\text{KK}}&=\cC\sum_{i=1}^3\sum_{n=1}^\infty|U_{\mrm{e}i}\cL^i_{0n}|^2\ep\sqrt{\ep^2-m_{i(n)}^2}\te(\ep-m_{i(n)}).
\end{align}
The kink nearest to $E_0$ is at $\ep=m_{i(0)}$. After this, in principle, all higher KK masses offer kinks at $\ep=m_{i(n)}$.
Since the minimum of $\ep$ is the sensitivity $0.4$\,eV, 
if some lighter modes are greater than but not too far from $0.4$\,eV, they can be detected by KATRIN.
For example, when $R=0.2\,\mu$m, $c_1=c_2=c_3=0$ and $\bar\mu_1=0.5$, the $\bt$-decay rate is given in Figure~\ref{sub_micron_KK}.
Here, for our convenience, we show only two kinks of the two lowest KK modes $n=1,2$ \footnote{See also \cite{Rodejohann:2014eka}.}. Note that not all KK modes can be included in the KATRIN observational window ($0-40~\text{eV}$). The largest number of $n_{\rm max}$ KK modes that KATRIN can detect is determined by the following condition:
\begin{align}
    m_{i(n_{\rm max})} \lesssim 40~{\rm eV}.
\end{align}
It is important to keep in mind that, as we discussed in Sec.~\ref{subsubsec:ubR-cot}, there is a theoretical upper bound on the size of the extra dimension in the case of small bulk mass ($ \pi|\bar c_i| \ll 1$ for $i=1,2,3$): $ R \lesssim 2~{\rm \mu m}$.  We now focus on the large bulk mass scenario ($\pi|\bar c_i| >1$), for which there is no upper bound on $R$, in order to look for the signal of the dark dimension right-handed neutrino in the $ R \sim \cO(10)\,{\rm \mu m}$ region.

\paragraph{Large bulk mass.}
Let us first concentrate on NH. 
We show that there is a parameter region 
where the contributions to the $\bt$-decay spectrum from infinitely many higher KK modes
can be described effectively as a contribution from a single mass with an effective mixing, 
so that the spectrum takes a similar form as \eqref{3+1_decay_rate} in the $(3+1)$-model.

\begin{figure*}
\begin{center}
  \includegraphics[width=1.0 \linewidth]{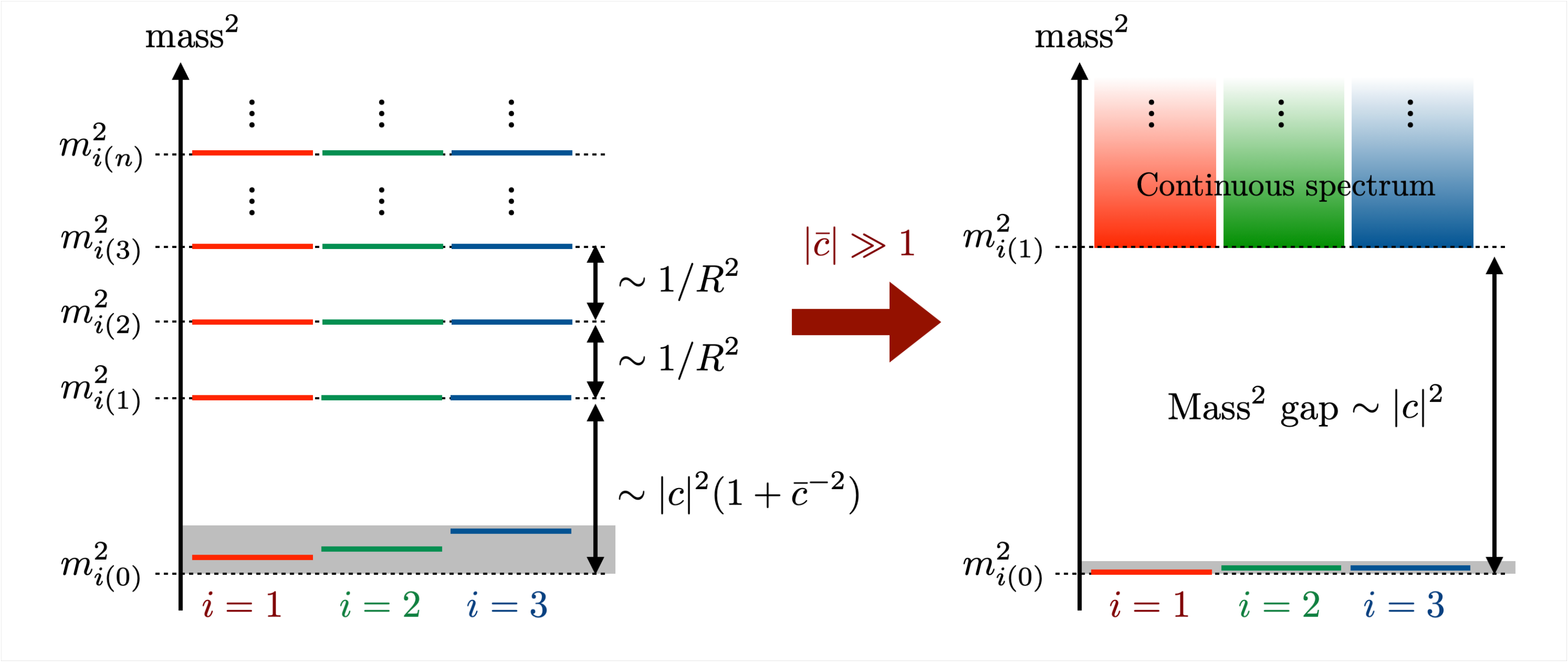}
 \caption{The left panel depicts a general structure of the mass spectrum. The right panel illustrates a behaviour in the large $|\bar c$ limit, where a mass gap between neighbouring KK modes becomes much smaller than the gap for the lightest two modes $n=0,1$. The density of the colour in each continuous KK mass spectrum depicts the magnitude of the mixing $\cL^i_{0n}$; the thicker the colour is, the larger the mixing is.}
 \label{KK_spectrum}
\end{center}
\end{figure*}

When $Rm_{i(0)}$ is suppressed, there is a squared mass gap between $n=0$ and $n=1$ of order $\de_1\sim|c_i|^2$ according to \eqref{approx-cothroot} and \eqref{al_i(n)} since $|\bar c_i|$ is large.
On the other hand, the squared mass gap between neighbouring higher KK masses is $\de_{\mrm{KK}}\sim R^{-2}$.
Therefore, the ratio $\de_{\mrm{KK}}/\de_1$ is given by $|\bar c_i|^{-2}$.
This indicates that in the large $|\bar c_i|$ limit, the ratio $\de_{\mrm{KK}}/\de_1$ approaches zero 
and the spectrum of higher KK modes becomes quasi-continuous compared with the gap between the zero and first modes.
The mixing with higher KK modes $\cL^i_{0n}$ ($n\geq1$) then has a peak around $n\sim|\bar c_i|$ and it decays for larger $n$.
Since the KK mass at the peak is $\sim\sqrt{2}|c_i|$, 
the main contribution to the higher KK part of the spectrum $\cR_{\mrm{KK}}$ is from the lighter region of the (almost) continuous KK spectrum of mass $\sim|c_i|$. 
This scenario is depicted in Figure~\ref{KK_spectrum}.

This argument can be quantified in the following parameter region:
\begin{align}
    |\bar c_i|\gg\max\{1,\,\bar\mu_i,\,\pi\bar\mu_i^2\},
    \label{approx-cond-2}
\end{align}
which has already been introduced in \eqref{approx-cond}.
Since $|\bar c_i|\gg1$, the zero mode masses solve the coth mass equations as proven in Sec.~\ref{subsec:mass-structure}.
Therefore, the zero mode masses $m_{i(0)}$ are given by \eqref{approx-cothroot}
and they are related by \eqref{mi2_Dem2}.
On top of the condition \eqref{approx-cond-2}, if the three bulk masses have similar values
\begin{align}
    |c_1| \sim |c_2| \sim |c_3|,
\end{align}
then the zero mode and the first KK mode $n=1$ have gap $|c_i|$, 
and the KK modes from $n=1$ to $n\sim|\bar c_i|$ will be of order $|c_i|$ for $i=1,2,3$.

Requiring that the first KK mass is above the sensitivity $|c_i|>0.4$\,eV
and taking the fact that maximum of the energy window of KATRIN is $40$\,eV into account, 
we can explore energies, for example, $0.5\,\text{eV}<|c_i|<40\,\text{eV}$, 
which for $R=10\,\mu\text{m}$ corresponds to $25<|\bar c_i|<2000$.
This is consistent with the assumption on the parameters \eqref{approx-cond-2} as long as $\bar\mu_1$ is taken appropriately.

For simplicity, we constrain the bulk masses to 
\begin{align}
    |c|:=|c_1|=|c_2|=|c_3|.
\end{align}
As mentioned above, since $|c|>0.4$\,eV and we are interested in large radius of around $10\,\mu$m, 
the three zero mode masses solve the coth mass equations, 
and the bulk masses for $i=2$ and 3 must be negative, as demonstrated in Sec.~\ref{subsubsec:ubR-coth}.
We may therefore look at the following two cases:
\begin{align}
    &c:=c_1=-c_2=-c_3>0, \label{c_positive}\\
    &c:=c_1=c_2=c_3<0, \label{c_negatice}
\end{align}
with $0.5\,\text{eV}<|c|<40\,\text{eV}$, and impose 
\begin{align}\label{cmu1-cond}
    |\bar c|\gg\max\{1,\,\bar\mu_i,\,\pi\bar\mu_i^2\}.
\end{align}
We can now approximate the spectrum from higher KK modes as follows.
It is convenient to rewrite the approximate formulas for $m_{i(n)},\cL^i_{0n}$ for $n\geq1$ given by \eqref{al_i(n)} and \eqref{Li0n} in terms of $n/|\bar c_i|$: 
\begin{align}
    (Rm_{i(n)})^2\simeq\bar c^2(1+n^2/\bar c^2), \qquad
    (\cL^i_{0n})^2\simeq\frac{2(\bar\mu_i^2/\bar c^2)(n^2/\bar c^2)}{(1+n^2/\bar c^2)^2},
\end{align}
Once the differential spectrum is rewritten in terms of $n/|\bar c|$, 
under \eqref{cmu1-cond}, we may replace $n/|\bar c|$ by a continuous variable $x$
as $\sum_nf(n/|\bar c|)/|\bar c|\to\int f(x)dx$
with the integration region from $1/|\bar c|$ to an upper bound $\Lm$
defined by $\ep\simeq m^i_{\Lm|\bar c|}$, as dictated by the step function $\te(\ep-m_{i(n)})$. 
Concretely, $\Lm$ reads
\begin{align}
    \Lm=\sqrt{\frac{(R\ep)^2}{\bar c^2}-1}=\sqrt{\frac{\ep^2}{c^2}-1}. \label{Lm-concrete}
\end{align}
Since $\Lm\geq1$, we need $\ep>|c|$.
Then the integral approximation of $\cR_{\text{KK}}$ is given by
\begin{align}
    \cR_{\text{KK}}
    &\simeq\cC\sum_{i=1}^3\sum_{n=1}^{\Lm|\bar c|}|U_{\mrm{e}i}|^2\frac{2(\bar\mu_i^2/\bar c^2)(n^2/\bar c^2)}{(1+n^2/\bar c^2)^2}\frac{\ep|\bar c|}{R}\sqrt{\frac{R^2\ep^2}{\bar c^2}-1-\frac{n^2}{\bar c^2}} \nn\\
    &\simeq\cC\sum_{i=1}^3 2\bar\mu_i^2|U_{\mrm{e}i}|^2\frac{\ep}{R}\int_{1/|\bar c|}^\Lm\frac{x^2}{(1+x^2)^2}\sqrt{\Lm^2-x^2}dx.
\end{align}
Dividing the integral region into $\int_0^\Lm-\int_0^{|\bar c|^{-1}}$ and taking the condition $\ep>|c|$ into account, we obtain\footnote{
    Concretely, we use
    \begin{align}
    &\int_0^\Lm\frac{x^2}{(1+x^2)^2}\sqrt{\Lm^2-x^2}dx=\frac{\pi(2+\Lm^2-2\sqrt{1+\Lm^2})}{4\sqrt{1+\Lm^2}}, \\
    &\int_0^{|\bar c|^{-1}}\frac{x^2}{(1+x^2)^2}\sqrt{\Lm^2-x^2}dx
    =\frac{|\bar c|^{-2}}{(1+|\bar c|^{-2})^2}\sqrt{\Lm^2-|\bar c|^{-2}}
    =\cO(|\bar c|^{-2}).
\end{align}
}
\begin{align}
    \cR_{\text{KK}}&\simeq
    \cC\left(\sum_{i=1}^3|U_{\mrm{e}i}|^2\frac{\pi\bar\mu_i^2}{2|\bar c|}\right)(\ep-|c|)^2\te(\ep-|c|),
    \label{sin2teeff}
\end{align}
where we used \eqref{Lm-concrete}.
This form makes it manifest that in the large $|\bar c|$ limit, 
$\cR_{\text{KK}}$ can be interpreted as a contribution from an effective mass $|c|$ with the effective mixing
\begin{align}
\sin^2\theta_{\text{eff}}:=\sum_{i=1}^3|U_{\mrm{e}i}|^2\frac{\pi\bar\mu_i^2}{2|\bar c|}.
\label{eff_mixing_def}
\end{align}
Note that the condition \eqref{cmu1-cond} guarantees that this effective mixing is smaller than 1.

Based on this result, let us rewrite the zero mode part $\cR_0$.
Using the approximate formula \eqref{Li00-approx} for $\cL^i_{00}$, 
we can rewrite the zero mode differential decay rate as
\begin{align}
    \cR_0
    &\simeq\cC\sum_{i=1}^3|U_{\mrm{e}i}|^2\left(1-\frac{\pi\bar\mu_i^2}{2|\bar c|}\right)
    \ep\sqrt{\ep^2-m_{i(0)}^2}\te(\ep-m_{i(0)}).
    \label{dGa0/dE-approx1}
\end{align}
As discussed for $\cR_{\mrm{std}}$, 
the active neutrino masses are degenerate and can be approximated by $m_\nu$ \eqref{m_nu_eff}.
Upon this replacement, \eqref{dGa0/dE-approx1} becomes
\begin{align}
    \cR_0\simeq(\cC\cos^2\theta_{\text{eff}})~\ep\sqrt{\ep^2-m_\nu^2}\,\te(\ep-m_\nu),
\end{align}
where we used the unitarity $1=\sum_{i=1}^3|U_{\mrm{e}i}|^2$ and the definition of $\sin^2\te_{\mrm{eff}}$ \eqref{eff_mixing_def}.
Finally, the total differential decay rate $\cR_\bt=\cR_0+\cR_{\mrm{KK}}$ can be rewritten as
\begin{align}
    \cR_\bt\simeq(\cC\cos^2\theta_{\text{eff}})~\ep\sqrt{\ep^2-m_\nu^2}\,\te(\ep-m_\nu)
    +(\cC\sin^2\theta_{\text{eff}})~(\ep-|c|)^2\te(\ep-|c|).
    \label{KK_decay_rate}
\end{align}
This has a quite similar form as the decay rate for the $(3+1)$ model \eqref{3+1_decay_rate}
 with the difference in the second term from sterile neutrinos.

Let us estimate the effective mixing $\sin^2\theta_{\text{eff}}$. 
Under \eqref{cmu1-cond}, $\bar\mu_i$ for $i=2,3$ can be approximated as
\begin{align}
    \bar\mu_i^2\simeq\bar\mu_1^2\te(-\bar c)+\frac{R^2\De m_{i1}^2}{2\pi|\bar c|}.
\end{align} 
The upper bound $R<30\,\,\mu\text m$ implies $R^2\De m_{21}^2 \ll R^2\De m_{31}^2<\cO(10)$. 
Since $|\bar c|\gg1$, we have ${R^2\De m_{i1}^2}/(2\pi|\bar c|)<\cO(1)$. Therefore, once \eqref{cmu1-cond} is satisfied for $\bar\mu_1$, it is satisfied automatically for $\bar\mu_2,\bar\mu_3$. 
According to neutrino oscillation experiments, we have $|U_{\mrm{e}1}|^2\sim0.96$ and $|U_{\mrm{e}i}|^2<0.04$. 
Therefore $|U_{\mrm{e}i}|^2{R^2\De m_{i1}^2}/(2\pi|\bar c|)$ is negligible. 
Therefore, $\sin^2\theta_{\text{eff}}$ can be approximated as
\begin{align}
    \sin^2\theta_{\text{eff}}
    \simeq\frac{\pi\bar\mu_1^2}{2|\bar c|}\left[|U_{\mrm{e}1}|^2+\te(-c)(|U_{\mrm{e}2}|^2+|U_{\mrm{e}3}|^2)\right]
    \simeq\frac{\pi\bar\mu_1^2}{|\bar c|}\times
    \begin{cases}
        0.48 & :c>0, \\
        0.5 & :c<0.
    \end{cases}
\end{align}
Since the setup assumes $\pi\bar\mu_1^2\ll|\bar c|$, the value of $\sin^2\theta_{\text{eff}}$ should be sufficiently smaller than 0.5.

\subsection{Comparison with 3$+$1 model}
Due to its ultra-luminous tritium source, KATRIN is well suited to search for sterile neutrinos with a mass $m_4$ in the range of eV to keV. Currently, the experiment measures the effective electron antineutrino mass by investigating the tritium $\beta$-decay spectrum in the last $40$ eV below the spectral endpoint at $E_0= 18.6$ keV, which is also the suitable energy range to search for eV-sterile neutrino \cite{KATRIN:2025lph}. The KATRIN neutrino mass program is planned to continue until the end of 2025. Afterwards, the beamline will be upgraded to enable a search for keV-scale sterile neutrinos.

In the previous section, we have seen the similarity between the $\bt$-spectra \eqref{3+1_decay_rate} and \eqref{KK_decay_rate}. In fact, in the parameter region \eqref{approx-cond-2} allowing the continuous limit, we could try to identify the mass of a sterile neutrino $m_4$ with the mass of the bulk fermion $|c|$ and also identify $\sin^2\theta_{ee}$ with the effective mixing $\sin^2\theta_{\rm eff}$ defined in \eqref{eff_mixing_def}. Under this identification, we could search for the existence of a dark dimension right-handed
neutrino by using KATRIN's sterile neutrino results. In this subsection, we explore this exciting possibility.

 \begin{figure*}
\begin{center}
  \includegraphics[width=0.55 \linewidth]{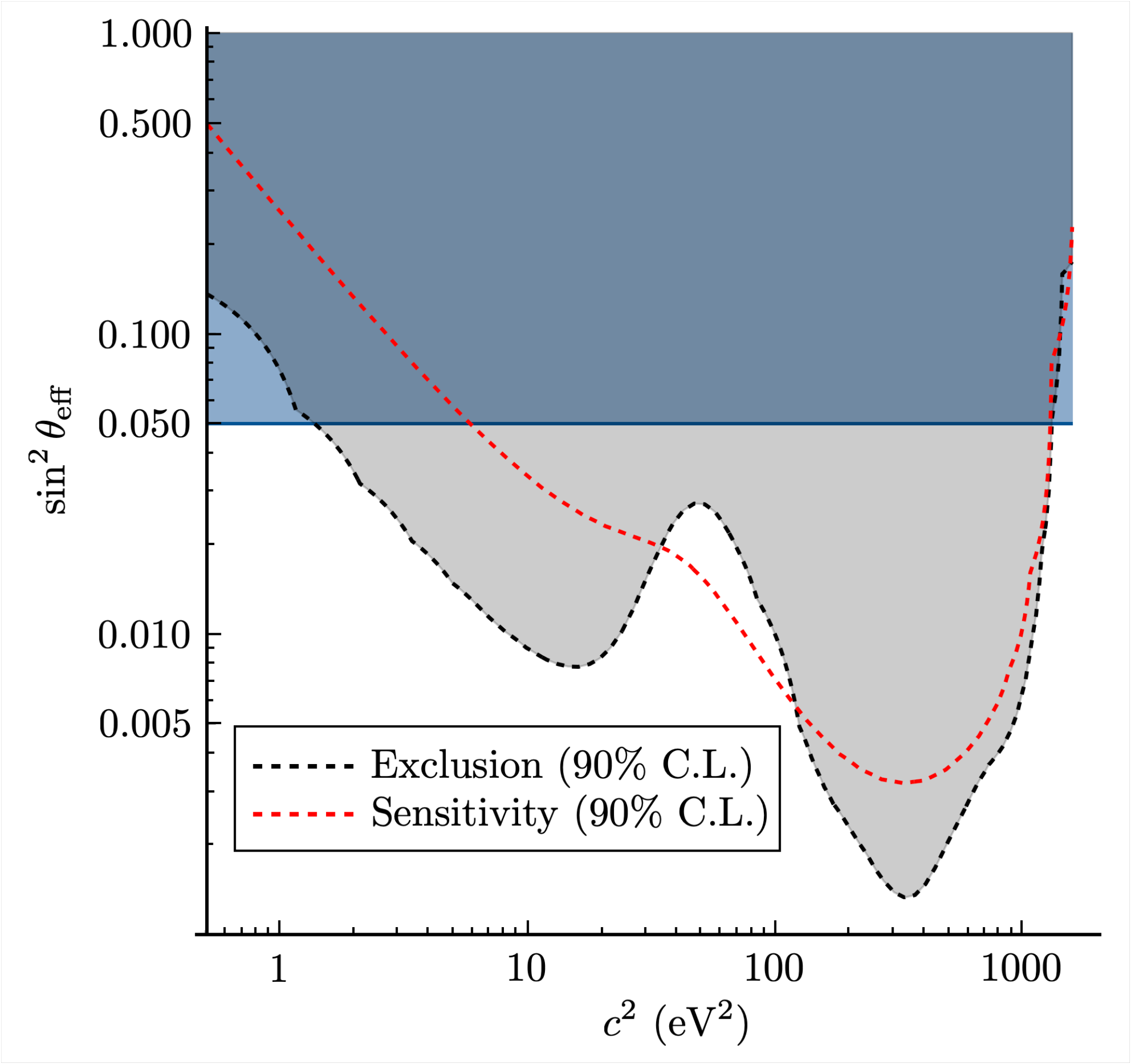}
 \caption{Sterile neutrino exclusion contour and the simulated sensitivity contour from KATRIN measurement campaigns KNM1--5 are shown in the black and red dashed line respectively \cite{KATRIN:2025lph}. The blue line represent the $\sin^2\theta_{\rm eff}^{\mrm{max}} = 0.050$ boundary. Below this upper bound, we can make the identification: $m_4^2 \rightarrow c^2$ and $\sin^2\theta_{ee} \rightarrow \sin^2\theta_{\rm eff}$.  }
 \label{3plus1exclusion}
\end{center}
\end{figure*} 

Let us discuss consequences from the cosmological bound on the active neutrino masses by Planck 2018 \eqref{Planck-ub}. 
In our regime \eqref{cmu1-cond}, the three zero mode masses are given by
\begin{align}
    m_{1(0)}&\simeq
    \begin{cases}
        \mu_1\sqrt{2\pi\bar c}e^{-\pi\bar c} & (c>0) \\
        \mu_1\sqrt{2\pi|\bar c|} & (c<0)
    \end{cases}, \qquad
    m_{i(0)}\simeq\sqrt{m_{1(0)}^2+\De m_{i0}^2}.
    \label{mi0-approx-equalc}
\end{align}
We can then see that 
if a given value of $\bar\mu_1$ is consistent with \eqref{Planck-ub}, 
then smaller values of $\bar\mu_1$ are also consistent.
Therefore, it makes sense to find, for each $\bar c$, the maximum value of $\bar\mu_1$ that is consistent with the cosmological bound.
Let us denote this maximum by $\bar\mu_1^{\text{max}}(\bar c)$. This gives the maximum of $\sin^2\te_{\text{eff}}$ for each $\bar c$,
\begin{align}
    \sin^2\te_{\text{eff}}^{\text{max}}(\bar c):=\frac{\pi\bar\mu_1^{\text{max}}(\bar c)^2}{2|\bar c|}.
    \label{sin2_max}
\end{align} 
In this way, we can make a curve $\sin^2\te_{\text{eff}}^{\text{max}}(\bar c)$ in $(c^2,\sin^2\te_{\mrm{eff}})$ plane for a given radius $R$,
and parameters below the curve are consistent with the Planck 2018 bound.

Note that $\bar\mu_1^{\text{max}}(\bar c)$ can be computed numerically independently of our approximation scheme \eqref{cmu1-cond}.
However, use of the effective mixing $\sin^2\te_{\text{eff}}^{\text{max}}(\bar c)$ only makes sense within this scheme. 
Therefore, when we make plots of $\sin^2\te_{\text{eff}}$, we have to specify a region where the approximation scheme \eqref{cmu1-cond} is valid.
Let us consider the cases $c_1>0$ and $c_1<0$ separately.
 \subsubsection{The negative $c$ case}
     Let us first consider the case in \eqref{c_negatice} with $c_1=c_2=c_3=-c<0$.
We impose the upper bound $m_{1(0)}<0.12\,$eV as implied by \eqref{Planck-ub}. Using \eqref{mi0-approx-equalc} for $m_{1(0)}$, we find
\begin{align}
    \bar\mu_1\sqrt{2\pi\bar c}<0.12\,(\text{eV} R)
\end{align} 
For example, when $R=10\,\mu$m, it gives an upper bound on $\bar\mu_1$,
\begin{align}
    \bar\mu_1<\frac{50.7\times0.12}{\sqrt{2\pi|\bar c|}}
    \label{mu1up-Planck}
\end{align}
The upper bound ranges from $0.44$ to $0.072$ as $\bar c$ ranges from $-25$ to $-2000$.
Therefore, the effective mixing $\sin^2\te_{\text{eff}}$ is suppressed to $<10^{-4}$. From Figure~\ref{3plus1exclusion}, this bound is much lower than the KATRIN sensitivity for $\sin^2\theta_{\rm eff}$.

We can impose another upper bound on the active Dirac neutrino mass coming from the swampland AdS conjecture that forbids stable non-supersymmetric AdS vacua~\cite{Ibanez:2017kvh, Anchordoqui:2023wkm}:
\begin{align}
    m_{1(0)}<7.63\,\text{meV},
    \label{swampAdS}
\end{align}
which yields a stronger upper bound on $\bar\mu_1$ than \eqref{mu1up-Planck}:
\begin{align}
    \bar\mu_1<\frac{50.7\times0.00763}{\sqrt{2\pi|\bar c|}}.
\end{align}

\subsubsection{The positive $c$ case}
 
Next, we consider the case $c_1=-c_2=-c_3=c>0$.
According to \eqref{mi0-approx-equalc}, $m_{1(0)}$ is exponentially suppressed to almost zero and hence $m_{i(0)}^2\sim\De m_{i1}^2$.
Therefore, the cosmological mass constraint \eqref{Planck-ub} is satisfied for any $R,c,\bar\mu_1$ as long as \eqref{cmu1-cond} is satisfied.\footnote{On the other hand, the mass bound by KATRIN gives $|U_{\text e1}|^2m_1^2+|U_{\text e2}|^2m_2^2+|U_{\text e1}|^2m_3^2<0.45^2\,\text{eV}^2$. Using $m_{i(0)}^2=m_{1(0)}^2+\De m_{i1}^2$ and the central values of the PMNS matrix components, one has $m_1<0.45\,\text{eV}$. In the case $c_1=c_2=c_3=-c<0$, for $R=10\,\mu$m, 
the upper bound on $\bar\mu_1$ ranges from $3.3$ to $0.54$ as $\bar c$ ranges from $-25$ to $-2000$. In the case $c_1=-c_2=-c_3=c>0$, the mass bound is trivially satisfied.}
Furthermore, the mass bound from the swampland AdS conjecture \eqref{swampAdS} is also satisfied for the same reason.

\begin{figure*}
\begin{center}
  \includegraphics[width=0.8 \linewidth]{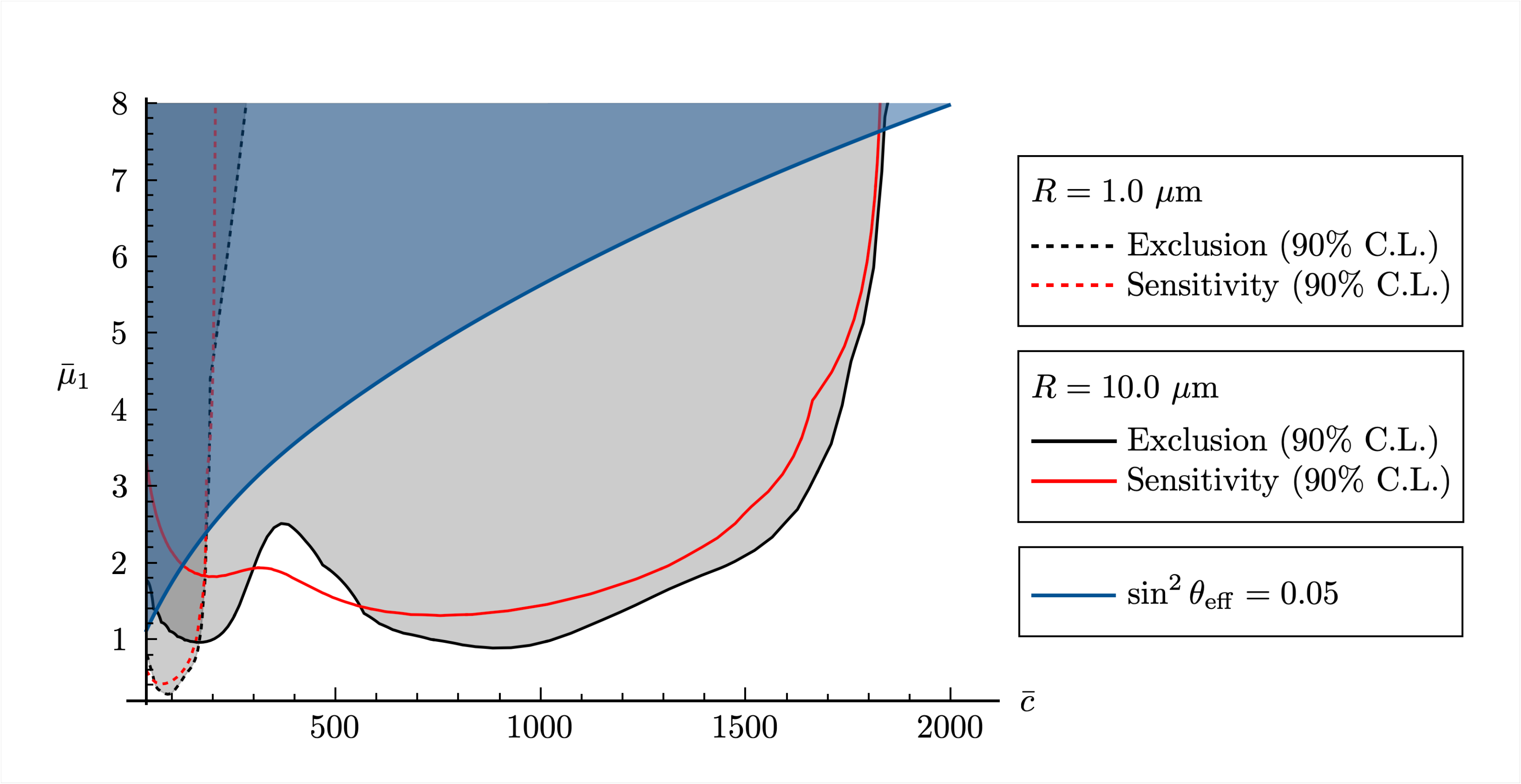}
 \caption{Exclusion contours for $R=1.0$ and $10.0~{\mu m}$ in the $(\bar c,\bar\mu_1)$-plane are shown as the black dashed and solid curves respectively. The blue curve represents the $\sin^2(\theta_{\rm eff}^{\mrm{max}}) = 0.050$ boundary.}
 \label{paraexclusion}
\end{center}
\end{figure*}

For demonstration purposes, we choose the maximum ratio to be $\pi\bar\mu_1^2/|\bar c| = 0.1$ to satisfy condition \eqref{approx-cond-2}. Equation \eqref{sin2_max} gives us the maximum value $\sin^2\theta_{\rm eff}^{\rm max} = 0.05$. In Figure \ref{3plus1exclusion}, the exclusion and sensitivity contours of the $3+1$ model are shown based on KATRIN's KNM1--5 measurement campaigns. Assuming that below the $\sin^2\theta_{\rm eff}^{\rm max} = 0.05$ boundary line, the decay rates in \eqref{3+1_decay_rate} and \eqref{KK_decay_rate} are indistinguishable, we can make the identification $m_4^2 \rightarrow c^2$ and $\sin^2\theta_{\mathrm{ee}} \rightarrow \sin^2\theta_{\rm eff}$. We can see that for the dark dimension neutrino with a mass parameter $|c|$ from $1.0$ to $10~{\rm{eV}}$, a large part of the parameter space is still allowed. The exclusion contours for $R= 1.0$ and $10.0~{\rm\mu m}$ in the $(\bar c,\bar\mu_1)$ plane are shown in Figure \ref{paraexclusion}.

Starting in 2026, KATRIN will be upgraded with the ``tritium investigation on sterile to active neutrino mixing" (TRISTAN) detector \cite{Siegmann:2024xvv}, enabling the measurements of the full tritrium $\beta$-decay spectrum. This allows the search for a sterile neutrino on a keV scale with an energy resolution of $300~{\rm eV}$. For dark dimension model with $R= 1.0-10.0\,{\mu}$m, TRISTAN can probe the region where $\bar c > 2000$.

\subsection{Comment on IH case}
In IH, we generate the mass spectrum by giving $\bar\mu_3$ by hand instead of $\bar\mu_1$. As in NH, we work with the equal bulk masses $|c|=|c_1|=|c_2|=|c_3|$. The parameter region is the same as \eqref{cmu1-cond}. Then the arguments leading to the effectively $3+1$ spectrum are not altered, and thus we can still use \eqref{KK_decay_rate} with the effective mixing \eqref{eff_mixing_def}.
For a given $\bar\mu_3$, by a parallel argument, the other two $\bar\mu_1^2,\bar\mu_2^2$ are given by
\begin{align}
    \bar\mu_i^2\simeq\bar\mu_1^2\te(-\bar c)+\frac{R^2\De m_{i3}^2}{2\pi|\bar c|}.
\end{align} 
Since ${R^2|\De m_{i3}^2|}/(2\pi|\bar c|)\lesssim\cO(1)$, all $\bar\mu_i^2$ satisfy \eqref{cmu1-cond}. The effective mixing $\sin^2\theta_{\mrm{eff}}$ can then be written as
\begin{align}
    \sin^2\theta_{\text{eff}}
    \simeq\frac{\pi\bar\mu_1^2}{2|\bar c|}\left[|U_{\mrm{e}3}|^2+\te(-c)(|U_{\mrm{e}1}|^2+|U_{\mrm{e}2}|^2)\right].
\end{align}
However, since $|U_{\mrm{e}1}|^2\sim0.96$ and $|U_{\mrm{e}i}|^2<0.04$, the effective mixing reads
\begin{align}
    \sin^2\theta_{\text{eff}}<\frac{\pi\bar\mu_1^2}{2|\bar c|}\times0.04,
    \label{sin2te-ub-IH}
\end{align}
whether $c$ is positive or negative. When $c<0$, even if a model with $(\bar c,\bar\mu_1=\bar\mu)$ in NH is above the sensitivity contour, an IH model with the same parameter set $(\bar c,\bar\mu_3=\bar\mu)$ can be much lower than the sensitivity contour in general.
In this case, $\bar\mu_3$ may have to be replaced by at least $\sim5\bar\mu$ to reach the sensitivity contour by increasing the upper bound in \eqref{sin2te-ub-IH} to ${\pi\bar\mu^2}/(2|\bar c|)$.

\section{Conclusion}

In conclusion, in this work we have made a detailed analysis of the experimental signatures of Right-handed neutrinos propagating along the ``Dark" Dimension of micron size that can be searched in KATRIN experiment, aiming to measure the value of the electron neutrino and look for possible existence of sterile neutrinos within the range of order 0.1--100 eV. The neutrino production should manifest as a kink in the differential beta decay spectrum as a function of the electron energy near the end point $E_0\simeq 18.57$ keV. Besides the size of the extra dimension $R$, every 5D R-neutrino brings two parameters: its bulk mass $c$ and its coupling to the SM brane $\mu$ defined in \eqref{mudef}, or equivalently the corresponding dimensionless parameters ${\bar c}=cR$ and ${\bar\mu =\mu R}$. We have identified two distinct regions in the parameter space, ${\bar c}\approx 0$ and ${\bar c}$ sufficiently large, as defined in \eqref{cmu1-cond}, where simple analytic expressions can be obtained, leading to qualitatively different experimental signatures within KATRIN's sensitivity. 

The first region leads to a series of kinks (in practice a few observable) corresponding to the production of KK excitations when $R$ is smaller than about half-micron and larger than a few nanometers. The upper bound is at the boarder of the limit obtained by the disappearance effect of neutrino oscillation data~\cite{Machado:2011jt, Forero:2022skg}. On the other hand, we found an allowed narrow window above this bound and below $\sim 2$ microns when $\bar\mu$ is large, for which the above limit does not apply due to the smallness of the active neutrino zero-mode coupling to the higher KK excitations. However this window falls slightly below the current sensitivity of KATRIN experiment.

The second region leads to effectively one kink around the value of the bulk mass due to an accumulation of KK excitations around this value with significant coupling to the active neutrino that can be parametrised by an effective mixing angle. The results in this region can thus be compared to those of the 3+1 sterile neutrino model that has been already analysed experimentally, leading to a constraint in the parameter space.

Our results are encouraging and motivate a dedicated numerical analysis in comparison with the experimental data in the whole parameter space of the model, which is currently ongoing with KATRIN collaboration.

\section*{Acknowledgements}

IA and HI are supported by the Second Century Fund (C2F), Chulalongkorn University. This research is funded by the NSRF via the Program Management Unit for Human Resources $\&$ Institutional Development, Research and Innovation (grant B39G670017) and by the Higher Education and Science Committee of RA (Research Project N 24RL-1C036). We are grateful to Joscha Lauer, Alexey Lokhov, Shailaja Mohanty, Jakkapat Seeyangnok, Narumon Suwonjandee from KATRIN collaboration for valuable discussions. We also thank Vichayanun Wachirapusitanand for assistance in the numerical analysis.

\appendix

\section{Technical details on our model of massive bulk neutrinos}
\label{app}

We first review the relation between the flavour and intermediate bases following~\cite{Machado:2011jt,Carena:2017qhd}.
The action in the intermediate basis \eqref{bulk_quad}$+$\eqref{bulkbrane_quad} is related to that in the flavour basis as
\begin{align}
    S_{\mrm b}&=\int\!d^4x\int_0^{\pi R}\!dz\,\Big(\msum_{\al}i\bar\Psi_\al\Ga^M\pd_M\Psi_\al-\msum_{\al,\bt}\cC_{\al\bt}\bar\Psi_\al\Psi_\bt\Big), \\
    S_{\mrm b\pd}&=-\Lambda_{\text{QG}}^{-1/2}\int\!d^4x\,\msum_{\al,\bt}\Big(\cY_{\al\bt}\ol{\ell_\al^{\text L}}\tilde H\Psi_\bt^\rR(z=0)+\text{c.c.}\Big),
\end{align}
where $\al,\bt$ run over the three favours $\mrm{e},\upmu,\uptau$.
The bulk-brane Yukawa matrix $\cY_{\al\bt}$ are related to the three Yukawa couplings $y_i$ in the interbediate basis via the singular value decomposition as
\begin{align}
    \cY_{\al\bt}=\msum_{i,j=1}^3U_{\al i}y_iV^\dg_{i\bt},
\end{align}
where $U$ and $V$ are $3\time3$ unitary matrices.
The bulk neutrinos and the SM leptons in the two bases are related by 
\begin{align}
    \Psi_\al=\msum_{i=1}^3V_{\al i}\Psi_i, \qquad
    \nu^\rL_\al=\msum_{i=1}^3U_{\al i}\nu^\rL_i, \qquad
    e_\al=\msum_{i=1}^3U_{\al i}e_i. 
\end{align}
which manifests that $U$ is the PMNS matrix.
The bulk mass matrix $\cC_{\al\bt}$ is then defined with the three bulk masses $c_i$ in the intermediate basis as
\begin{align}
    \cC_{\al\bt}=\msum_{i=1}^3V_{\al i}c_iV^\dg_{i\bt}.
\end{align}

Let us proceed to the diagonalisation of the mass matrix in \eqref{4dKKaction-Sb}$+$\eqref{bulkbrane_4D}.
It is convenient to rewrite the total quadratic action $S_{\mrm{tot}}:=S_{\mrm{b}\nu_\rL}+S_{\mrm{b}\pd}$ in the matrix notation with respect to the KK labels:
\begin{align}
S_{\mrm{tot}}=\int\!d^4x\,\msum_{i=1}^3\Big[i\ol{\bscN_i^{\rL}}\ol\sig{}^\mu\pd_\mu\bscN_i^{\text L}+i\ol{\bscN_i^{\rR}}\sig^\mu\pd_\mu\bscN_i^{\rR}-\ol{\bscN_i^{\rR}}\sfM_i^\dg\bscN_i^{\rL}-\ol{\bscN_i^{\rL}}\sfM_i\bscN_i^{\rR}\Big],
\end{align}
where the vectors $\bscN_i^\rL,\bscN_i^\rR$ and the mass matrix $\sfM_i$ are defined by
\begin{align}
\bscN_i^\rL=\begin{bmatrix} 
\nu^{\rL}_i \\  
\psi^{\rL}_{i1} \\
\psi^{\rL}_{i2} \\
\vdots \\
\psi^{\rL}_{in} \\
\vdots
\end{bmatrix}, \quad
\bscN^{\rR}_i=\begin{bmatrix} 
\psi^{\rR}_{i0} \\  
\psi^{\rR}_{i1} \\
\psi^{\rR}_{i2} \\
\vdots \\
\psi^{\rR}_{in} \\
\vdots
\end{bmatrix}, \quad
\sfM_i=\begin{bmatrix} 
Y_{i0} & Y_{i1} & Y_{i2} & \cdots & Y_n & \cdots \\  
0 & \lm_{i1} & 0 & \cdots & 0 & \cdots \\
0 & 0 & \lm_{i2} & \cdots & 0 & \cdots \\
\vdots & \vdots & \vdots & \ddots & \vdots & \vdots \\
0 & 0 & 0 & \cdots & \lm_{in} & \cdots \\
\vdots & \vdots & \vdots & \cdots & \vdots & \ddots
\end{bmatrix}.
\label{mass-matrix}
\end{align}
The mass term can be diagonalised through the singular value decomposition of $\sfM_i$: 
\begin{align}
\sfM_i=\sfL_i\sfD_i\sfR_i^\dg, \quad \sfM_i\sfM_i^\dg=\sfL_i\sfD_i^2\sfL_i^\dg,
\end{align}
where $\sfL_i,\sfR_i$ are unitary, 
and $\sfD_i$ is the diagonal matrix with its non-negative diagonal elements being the square root of the eigenvalues of $\sfM_i\sfM_i^\dg$.
Let us denote the diagonal elements of $\sfD_i$ by $\{m_{i(n)}:n=0,1,2,\cdots\}$ in the increasing order,
\begin{align}
m_{i(0)}<m_{i(1)}<m_{i(2)}<\cdots.
\end{align}
We can show, as demonstrated later, that they are given as the positive roots of the equation:
\begin{align}
h_i(Rm_{i(n)})=0,
\end{align}
where the function $h_i(x)$ is given as
\begin{align}\label{hi-def2}
h_i(x):=x^2 + \pi(Rc_i)(R\mu_i)^2 - \pi(R\mu_i)^2\sqrt{x^2-(Rc_i)^2}\cot\Big(\pi\sqrt{x^2-(Rc_i)^2}\Big).
\end{align}
After the singular value decomposition, the action is rewritten as
\begin{align}
S_{\mrm{b}\nu_\rL}+S_{\mrm{b}\pd}
&=\int\!d^4x\,\msum_{i=1}^3\Big[i\ol{\bsN_i}\ga^\mu\pd_\mu\bsN_i-\ol{\bsN_i}\sfD_i\bsN_i\Big],
\end{align}
where we introduced the vectors of fields $\bsN_i^\rL,\bsN_i^\rR,\bsN_i$ by
\begin{align}
\bscN_i^\rL=\sfL_i\bsN_i^\rL, \quad \bscN_i^\rR=\sfR_i\bsN_i^\rR, \quad \bsN_i=\bsN_i^\rL+\bsN_i^\rR.
\end{align}
For concreteness, we introduce the component notations:
\begin{align}
&\nu_{i(\ell)}^\rL:=(\bsN_i^\rL)_\ell, \quad \nu_{i(\ell)}^\rR:=(\bsN_i^\rR)_\ell, \quad 
\nu_{i(\ell)}:=\nu_{i(\ell)}^\rL+\nu_{i(\ell)}^\rR \quad (\ell\geq 0), \\
&\cL^i_{n\ell}:=(\sfL_i)_{n\ell}, \quad \cR^i_{n\ell}:=(\sfR_i)_{n\ell}, \quad (n,\ell\geq 0), \\
&m_{i(\ell)}=(\sfD_i)_{\ell\ell}, \quad m_{i(0)}<m_{i(1)}<m_{i(2)}<\cdots \quad (\ell\geq 0). \label{mia}
\end{align}
In terms of them, the action is given by \eqref{nus-massbasis}.

Let us demonstrate the diagonalisation of the matrix $\sfM_i\sfM_i^\dg$ given by
\begin{align}
\sfM_i\sfM_i^\dg=
\begin{bmatrix} 
\sum_{n=0}^\infty|Y_{in}|^2 & (Y_{i1})(\lm_{i1}) & (Y_{i2})(\lm_{i2}) & \cdots & (Y_{iK})(\lm_{iK}) & \cdots \\  
(Y_{i1}^*)(\lm_{i1}) & (\lm_{i1})^2 & 0 & \cdots & 0 & \cdots \\
(Y_{i2}^*)(\lm_{i2}) & 0 & (\lm_{i2})^2 & \cdots & 0 & \cdots \\
\vdots & \vdots & \vdots & \ddots & \vdots & \vdots \\
(Y_{iK}^*)(\lm_{iK}) & 0 & 0 & \cdots & (\lm_{iK})^2 & \cdots \\
\vdots & \vdots & \vdots & \cdots & \vdots & \ddots
\end{bmatrix}.
\end{align}
The determinant $\det(R^2\sfM_i\sfM_i^\dg-x^2I)$ $(x\geq 0)$ is then given by
\begin{align}
&\det(R^2\sfM_i\sfM_i^\dg-x^2I)
=\lim_{K\to\infty}\prod_{k=1}^K\big[(R\lm_{ik})^2-x^2\big]
\left(\sum_{n=0}^K|RY_{in}|^2-x^2-\sum_{n=1}^K\frac{|RY_{in}|^2(R\lm_{in})^2}{(R\lm_{in})^2-x^2}\right) \nn\\
&\qquad\qquad 
=\lim_{K\to\infty}\prod_{k=1}^K\big[(R\lm_{ik})^2-x^2\big]
\left(|RY_{i0}|^2-x^2-x^2\sum_{n=1}^K\frac{|RY_{in}|^2}{(R\lm_{in})^2-x^2}\right).
\end{align}
When $x\sim R\lm_{in}$, the RHS becomes $-(R\lm_{in})^2|RY_{in}|^2\prod_{k\neq n}[(R\lm_{ik})^2-(R\lm_{in})^2]\neq0$. 
Therefore, the characteristic equation $\det(R^2\sfM_i\sfM_i^\dg-xI)=0$ is equivalent to the equation
\begin{align}\label{KKmasseq1}
|RY_{i0}|^2-x^2-x^2\sum_{n=1}^\infty\frac{|RY_{in}|^2}{(R\lm_{in})^2-x^2}=0.
\end{align}
By using \eqref{Y0i}, this equation can be rewritten into \eqref{transeq1} 
(and also \eqref{transeq2} for $x < |\bar c_i|$). 

Let us proceed to $\cL^i_{n\ell}$. Recall that $\sfM_i\sfM_i^\dg=\sfL_i\sfD_i^2\sfL_i^\dg$. 
Let us denote $\sfL_i=(\bsv^i_0,\bsv^i_1\,\bsv^i_2\,\cdots)$ where the vector $\bsv^i_\ell$ is defined by $(\bsv^i_\ell)_n:=\cL^i_{n\ell}$. 
Then $\sfM_i\sfM_i^\dg\sfL_i=\sfL_i\sfD_i^2$ is equivalent to the eigenvalue equations
\begin{align}
\sfM_i\sfM_i^\dg\bsv^i_\ell=m_{i(\ell)}^2\bsv^i_\ell.
\end{align}
Since $\sfL_i$ is unitary, $\bsv^i_{\ell}$ has norm 1. 
Solving this equation yields \eqref{mixmat-0a} and \eqref{mixmat-na}.
The other unitary matrix $\cR^i_{n\ell}$ can be obtained in a parallel manner by diagonalising $\sfM_i^\dg\sfM_i$.


\providecommand{\href}[2]{#2}\begingroup\raggedright\endgroup

\end{document}